\documentclass[nofootinbib,Ffigurea4paper,fleqn,aps,prd,10pt,superscriptaddress,reprint,showkeys,Fdelt_cshowpacs]{revtex4-1}
\usepackage[title,titletoc]{appendix}
\usepackage{subfig}
\usepackage{times}
\usepackage{amsfonts}
\usepackage{amsmath}
\usepackage{amssymb}
\usepackage{natbib}
\usepackage{graphicx}
\usepackage{enumitem}
\usepackage{xcolor}
\usepackage[colorlinks=true,linkcolor=blue]{hyperref}
\usepackage[outdir=./]{epstopdf}
\usepackage[normalem]{ulem}

\def\aap{A\&A}
\def\aj{AJ}
\def\apj{ApJ}
\def\apjl{ApJL}
\def\apjs{ApJS}
\def\apss{Astrophysics and Space Science}
\def\jcap{JCAP}
\def\mnras{MNRAS}
\def\na{New Astronomy}

\def\physrep{Physics Reports}

\begin{document}

\title{\bf Shear, vorticity, and $f(R)$ theories
}

\author{Antonino~\surname{Del Popolo}}
\affiliation{Dipartimento di Fisica e Astronomia, University of Catania,
Viale Andrea Doria 6, 95125, Catania, Italy}
\affiliation{Perimeter Institute, 31 Caroline Street North, Waterloo,
Ontario, N2L 2Y5 Canada}

\label{firstpage}

\date{\today}

\begin{abstract}
In this paper, we study how an extended spherical collapse model (ESCM) taking into account the interaction with external protostructures through shear and vorticity, and screening, modifies typical parameters of the collapse, such as: the collapse threshold, $\delta_c(z)$ and $\delta_c(M)$, the turnaround density, $\delta_t$, and the turnaround radius, $R_{\rm t}$, and halo mass function (HMF). Differently from the standard spherical collapse model  (SSCM) which excludes shear and vorticity (quantities depending on mass), the collapse threshold $\delta_c$ depends on mass, redshift, and the parameter $|f_{\rm R0}|$ of the Hu-Sawicki $f_{\rm R}$ theory. 
We used this result to obtain the halo mass function by means of the excursion set approach. The main goal of the paper is to understand if the quoted quantities, and the mass function can disentangle between General Relativity (GR), and the Hu-Sawicki $f(R)$ model. When considering the ESCM prediction for GR and the large field (LF) limit of the $f(R)$ theory, the percentage difference between $\delta_c$ is of a few percent and is around $20 \%$ if we consider the difference between the GR in the ESCM, and the LF limit in the SSCM. In the case of $\delta_t$, the values are around $27$ and $29 \%$, while for $R_t$ is around $6.5 \%$. 
In the case of $\delta_c$ and for $|f_{R0}|=10^{-4}$ (F4) the difference between the GR value of $\delta_c$, and the screened one is $2.5 \%$, and goes down to 0 for $|f_{R0}|=10^{-6}$ (F6). In the case of $\delta_t$ they are larger than $20 \%$ for F4, $\simeq 10 \%$ for F5, and $0 \%$ for F6. In the case of the turnaround radius $\simeq 6.5 \%$ (F4), $\simeq 3 \%$ (F5), and $\simeq 0 \%$ (F6). The observational constraints on $|f_{R0}|$ give values smaller than F4 ($|f_{R0}| \lesssim 10^{-4.5}$, at large scales), thus only F5 could be used to disentangle between GR and $f(R)$, since F4 is excluded, $F6$ is practically equal to GR, and the LF is excluded because it is larger than F4. Unfortunately, as shown by some studies, the error on $R_t$, is of the order of $20\%$, excluding the possibility to use the turnaround radius, and related quantities, to disentangle between GR and $f(R)$. 

Concerning the mass function, the haloes number predicted by $f(R)$, especially at high redshift can disentangle between $f(R)$ and GR. 

\end{abstract}



\maketitle

\section{Introduction}

The $\Lambda$CDM model, based on the assumption that GR is the correct theory of gravitation, has successfully passed many tests \citep{Komatsu2011,Planck2014_XVI,DelPopolo2013}. Nowadays it has risen to the position of "the standard model of cosmology". The $\Lambda$CDM model is based on the idea that the Universe contains a larger content of mass-energy than predicted \cite{DelPopolo2007,DelPopolo2014,Bull2016}, dominated by non-baryonic and non-relativistic particles, 
indicated as "cold dark matter" \cite{DelPopolo2014,Arbey2021}. 
Unfortunately, DM particles, despite a large campaign of direct and indirect searches \cite{Bertone2005,Klasen2015, DelPopolo2014}, and indirect evidences from small to large scales  \cite{Einasto2001,Bertone2005,Bouchet2004,Kilbinger2015}, have never been observed \cite{Klasen2015}. XENON1T \citep{Aprile2018} has not found 
significant excess over the background, in its direct-detection experiments, and the same happened with other direct-detection experiments. Indirect searches have also been unsuccessful. A large parameter space for annihilating dark matter was ruled out based on studies of the Andromeda galaxy \citep{Chan2019} and A2877 and Fornax clusters \citep{Chan2017}.
A second fundamental component of the standard cosmology, is the so-called ``dark energy'' (DE), a fluid having negative pressure, producing an accelerated expansion of the Universe \cite{Riess1998,Perlmutter1999}. In its simplest form, and in the $\Lambda$CDM model, DE is represented by the cosmological constant $\Lambda$. Unfortunately, the cosmological constant $\Lambda$ 
suffers from the ``cosmological constant fine tuning problem", and the ``cosmic  coincidence problem" \cite{Astashenok2012,Velten2014,Weinberg1989}.

Although the $\Lambda$CDM model is celebrated as a great success of science \cite{Spergel2003,Komatsu2011,DelPopolo2007}, several drawbacks are being discovered at small scales \cite{Moore1999,deBlok2010,Ostriker2003,BoylanKolchin2011,DelPopolo2014a,DelPopolo2014d,DelPopolo2017a}, and large ones 
\citep{Eriksen2004,Schwarz2004,Cruz2005,Copi2006,Macaulay2013,Planck2014_XVI,Raveri2016}.

The issues at small scale are usually called "small scale problems" of the $\Lambda$CDM \cite{DelPopolo2017a}. 
Several recipes, such as astrophysical ones \cite{Brooks2013,Onorbe2015,DelPopolo2017a}, different nature of the dark matter particles \citep{Colin2000,Goodman2000,Hu2000,Kaplinghat2000,Peebles2000,SommerLarsen2001}, or modified gravity (MG) theories, the simplest being the Modified Newton Dynamics (MOND) \citep{Milgrom1983} model, $f(R)$ \citep{Buchdahl1970,Starobinsky1980}, $f(T)$ \citep{Bengochea2009,Linder2010,Dent2011,Zheng2011} have been proposed to solve those problems.

The issues related to the cosmological constant $\Lambda$ motivated researchers to look for other explanations, and models to clarify the universe accelerated expansion. DE effects have been explained through additional matter fields (e.g., quintessence \citep{Copeland2006}), or again with MG models (e.g., $f(R)$ theories)\citep{Horndeski1974,Milgrom1983,Zwiebach1985,Moffat2006,Nojiri2005,Bekenstein2010,DeFelice2010,Linder2010,
Milgrom2014,Lovelock1971,Horava2009,Rodriguez2017,Deffayet2010}. Accelerated expansion explanation, has brought on the table  extra dimensions or higher-order corrections effects, as in the Dvali-Gabadadze-Porrati (DGP) model \cite{Dvali2000}.

To understand in which direction to move to solve the quoted problems is not easy. A solution, or at least a better understanding of the problems could come from new studies of the CMB \citep{Battye2018a,Battye2018c}, or present or future surveys like SKA, LSST, EUCLID, JDEM, etc. 

MG effects can be disentangled from GR by means of scale relationships in clusters like the mass-temperature relation (MTR) \cite{Hammami2017,DelPopolo2019}, the luminosity-temperature relation (LTR) \cite{DelPopolo2026}, 
or by structure formation. Important probes are the splashback radius \cite{Adhikari2018}, and the turnaround radius (TAR) \cite{Pavlidou2014,Pavlidou2014a,Faraoni2015,Bhattacharya2017,Lopes2018,Lopes2019,DelPopolo2020a,DelPopolo2020b}.  

TAR\footnote{Defined as the distance from the center of a structure to the surface at which radial velocity is zero. The majority of works studies the maximum TAR, defined as the radius at which the radial acceleration is null, $\ddot{r} =0$.} has been proposed as a good way to test cosmological models \citep{Lopes2018}, DE, and disentangle between $\Lambda$CDM model, DE, and MG models \citep{Pavlidou2014,Pavlidou2014a,Faraoni2015,Bhattacharya2017,Lopes2018,Lopes2019,DelPopolo2020a,DelPopolo2020b}.

When \citep{Lopes2019} showed that MG can affect the maximum TAR, the attention on this quantity increased. Another reason the TAR has been studied is because according to several authors it is well defined and not influenced by baryonic physics (e.g., \citep{Pavlidou2014}). 

This point is not exactly correct since as shown by \citep{DelPopolo2013a,DelPopolo2013b,Pace2014,Mehrabi2017,Pace2019,DelPopolo2020a,DelPopolo2020b} the 
non-linear equation describing the evolution of the overdensity contrast depends on several quantities such as dynamical friction, shear and vorticity \citep{DelPopolo2020a} which depends on the structure mass, and baryonic physics. In the majority of papers dealing with the study of TAR the real physics of structure formation is not taken into account.

In the present paper, we use an extended spherical collapse model (ESCM) introduced, and adopted in   \citep{DelPopolo2013,DelPopolo2013a,Pace2014,Mehrabi2017,Pace2019}, taking into account the effect of a non null shear and vorticity on the collapse. 
As we have already shown, shear and vorticity change the typical parameters of the SSCM \citep{DelPopolo2013}, the mass function \citep{DelPopolo2013,DelPopolo2013a,Pace2014,Mehrabi2017}, the two-point correlation function \citep{DelPopolo1999}, and the weak lensing peaks \citep{Pace2019}. 
Here, we added the Chameleon mechanism in a Hu-Sawicki $f(R)$ theory. 

$f(R)$ models are a class of models replacing in Einstein-Hilbert action the Ricci scalar curvature $R$ by $R + f(R)$ studied in many contexts. In $f(R)$ theory, \citep{Capozziello2019} found a general relation for the maximum TAR, \citep{Faraoni2015} generalized the result to generic gravitational theories\footnote{\citep{DelPopolo2020a} found a more general formula than that of \citep{Pavlidou2014} for smooth DE, showing that shear, and vorticity, modify TAR.}.

In the paper we use the Hu-Sawicki model \citep{Sawicki2007,Hu2007} and use our extended spherical collapse model (ESCM). 
SSCM has been applied to $f(R)$ theories in several papers (e.g., \citep{Borisov2012,Kopp2013}). In these papers, SSCM has been used to determine the collapse threshold, and the halo mass function. The goal of the present paper is to see how the collapse parameters, $\delta_c$, $\delta_t$, the turnaround radius, and the mass function are modified. We will also show that the collapse threshold depends on mass, redshift, and $f_{\rm R0}$. This threshold, as in \citep{DelPopolo1998,DelPopolo1999,DelPopolo2000, DelPopolo2006c,DelPopolo2013,DelPopolo2013a,Pace2014,Mehrabi2017,Pace2019} being not a constant as in the EdS Universe, or approximately constant in the $\Lambda$CDM, furnishes the barrier needed in the excursion set formalism to get the multiplicity function and mass function (see also \citep{SMT2001}), that we obtained. 

We want to discuss if the turnaround radius $R_t$ is really a tool to test MG models \citep{Lee2016,Pavlidou2014,Pavlidou2014a,Tanoglidis2015,Tanoglidis2016}, and if the mass function is a better probe. 

In this paper, we try to put constraints on $f_{\rm R0}$ values by means of turnaround, and the mass function obtained starting from the threshold of collapse, $\delta_c(M)$, that, apart from a numerical normalization constant is the "collapse barrier" used in the excursion set approach to obtain the mass function.

The paper is organized as follows: 
Section~\ref{sect:model} describes the model used. 
Section~\ref{sect:Results} is devoted to the presentation and the discussion of our results.  
Section~\ref{sect:conclusions} is devoted to conclusions.

\section{The Model}\label{sect:model}

In 1972, Gunn \& Gott \cite{Gunn1972} introduced a very simple SCM model just considering the radial collapse of a structure. 
Despite its simplicity it is a very popular method to study analytically the non-linear evolution of perturbations of DM and DE, etc. Their SCM does not take into account random angular momentum  \citep{Ryden1987,Ryden1988,Williams2004,Fillmore1984,Bertschinger1985,Hoffman1985,Ryden1987,Subramanian2000,
Ascasibar2004,Williams2004}, tidal angular momentum \citep{Peebles1969,White1984}, dynamical friction (\citep{AntonuccioDelogu1994,Delpopolo2009}), etc. The way to take into account angular momentum into account
was found by \citep{Ryden1987,Gurevich1988a,Gurevich1988b,White1992,Sikivie1997,Nusser2001,Hiotelis2002,  
LeDelliou2003,Ascasibar2004,Williams2004,Zukin2010}\footnote{In order for the mean angular momentum at any point in space to be zero \cite{White1992,Nusser2001}, particles' angular momentum is distributed in random directions}, that of dynamical friction was studied in \citep{AntonuccioDelogu1994,Delpopolo2009}, while \citep{Hoffman1986,Hoffman1989,Zaroubi1993} discussed the role of shear in the gravitational collapse.
The SSCM with negligible DE perturbations was studied by \citep[see, e.g.][]{Bernardeau1994,Bardeen1986,Ohta2003,Ohta2004,Basilakos2009,Pace2010,Basilakos2010} 
while in \citep[see][]{Mota2004,Nunes2006,Abramo2007,Abramo2008,Abramo2009a,Abramo2009b,Creminelli2010,Basse2011,Batista2013}
the effects of the DE fluid perturbation were taken into account.  
\citep{Pettorino2008,Wintergerst2010a,Tarrant2012} extended the SSCM model to coupled DE models, and 
\citep{Pettorino2008,Pace2014} to extended DE (scalar-tensor) models. \cite{DelPopolo2013a,DelPopolo2013b} studied the effects of shear and rotation in smooth DE models. \cite{Pace2014} studied SSCM in clustering DE cosmologies, and 
\cite{DelPopolo2013c} in Chaplygin cosmologies. As reported in the introduction, it was used also in MG theories. 

In this paper, we use the ESCM taking into account shear and vorticity, and chameleon effect 
in the case of the Hu-Sawicki model. In order not to have a too long section concerning the model, we wrote the model's details in some appendices. Appendix A deals with the $f(R)$ models and Hu-Sawicki model. Appendices B, and C, discuss how the evolution equation for $\delta$, with shear and vorticity, was obtained. Appendix D, discusses the chameleon mechanism and the way it is introduced in the model.  
Appendix E
discusses the goodness of the quasi-spherical approximation, already shown in other papers of ours. 

As shown in Appendix C, the equation that we need to solve to determine the properties of spherical collapse in Hu-Sawicki model is Eq. \ref{eq:HuSaw}, an integro-differential equation, together with its linearized form. 
Eq. (\ref{eq:HuSaw}) must also be modified to take the chameleon effect and screening into account.
This issue is discussed in Appendix D. 

The evolution equation, with shear, vorticity, and screening (using the thin-shell method) is given by: 
\begin{equation} \label{eq:HuSaw}
\begin{aligned}
&\delta''(r_i) + \left(\frac{3}{a}+\frac{E'}{E}\right)\delta'(r_i) -\frac{4}{3}\frac{\delta'(r_i)^2}{1+\delta(r_i)} = \\
&\frac{3(1+\delta(r_i))\Omega_{m,0}}{2a^5E^2} \left[\mathcal{G}(r_i) - \alpha\,\delta(r_i)\right]
\end{aligned}
\end{equation}
(see Eq. \ref{eq:mast})
where 
\begin{equation}
\mathcal{G}(r_i) = \frac{1}{2\pi^2}\int_0^\infty dk\, k^2\, \big[1+S(r_i,a)\,\epsilon(k,a)\big]\,
\delta(k,a)\,\frac{\sin(kr_i)}{kr_i}.
\label{eq:mast1}
\end{equation}

where $S(R_i,a)$ is the Khoury-Weltman condition given in Eq. \ref{eq:KW}, and $\epsilon(k,a)$ is given by

\begin{equation} \label{eq:eps}
\epsilon(k,a)
=
\frac{1}{3}
\frac{k^2}{a^2m_{f_R}^2+k^2}.
\end{equation}

where $' = \frac{d}{da}$, and $E$ is given by Eq. \ref{eq:e}. 

To study the nonlinear evolution of the matter density contrast, an initial overdensity
profile, $\delta_i(r)$, must first be specified. In $f(R)$ theories, the shape of an initial top-hat density 
profile, will evolve, developing a large spike close to the boundary between the overdensity 
and background FRW spacetime. 

This profile is evolved from an early epoch, which we choose to be $z_i=500$, where the effects of modified gravity are still negligible and the evolution is practically indistinguishable from that predicted by GR.
Following the approach adopted in previous studies \textbf{\cite{Kopp2013,Lopes2018}}, we consider an initial density
profile referred to as the {Tanh} profile characterized by the size of the top-hat-like function $r_b$ denoting the characteristic radius of the overdense region and 
$s$ controlling the sharpness of the profile boundary (the steepness of the transition). In the limit $s\rightarrow0$, the profile approaches the ideal top-hat configuration:
\begin{equation}
\delta_i(r)=
\frac{\delta_{i,0}}{2}
\left[
1-
\tanh\left(
\frac{r/r_b-1}{s}
\right)
\right],
\label{eq:tanh_profile}
\end{equation}

Since the contribution of $\delta(r)$ to the mean initial density of the spherical region is negligible, at the initial time
the enclosed mass can be approximated by $M_b\simeq
\frac{4\pi}{3}\,\bar{\rho}_0\,r_b^3$. 
Since the mass enclosed within the spherical perturbation remains constant, $M=\frac{4\pi}{3}\,\bar{\rho}\,r^3(1+\delta)$,
the radius of the overdense region is directly related to the nonlinear density contrast, $r^{3}\,\bar{\rho}\,(1+\delta)=\frac{3M}{4\pi}$.

\begin{figure*}[htbp]
\centering

\includegraphics[width=0.9\textwidth]{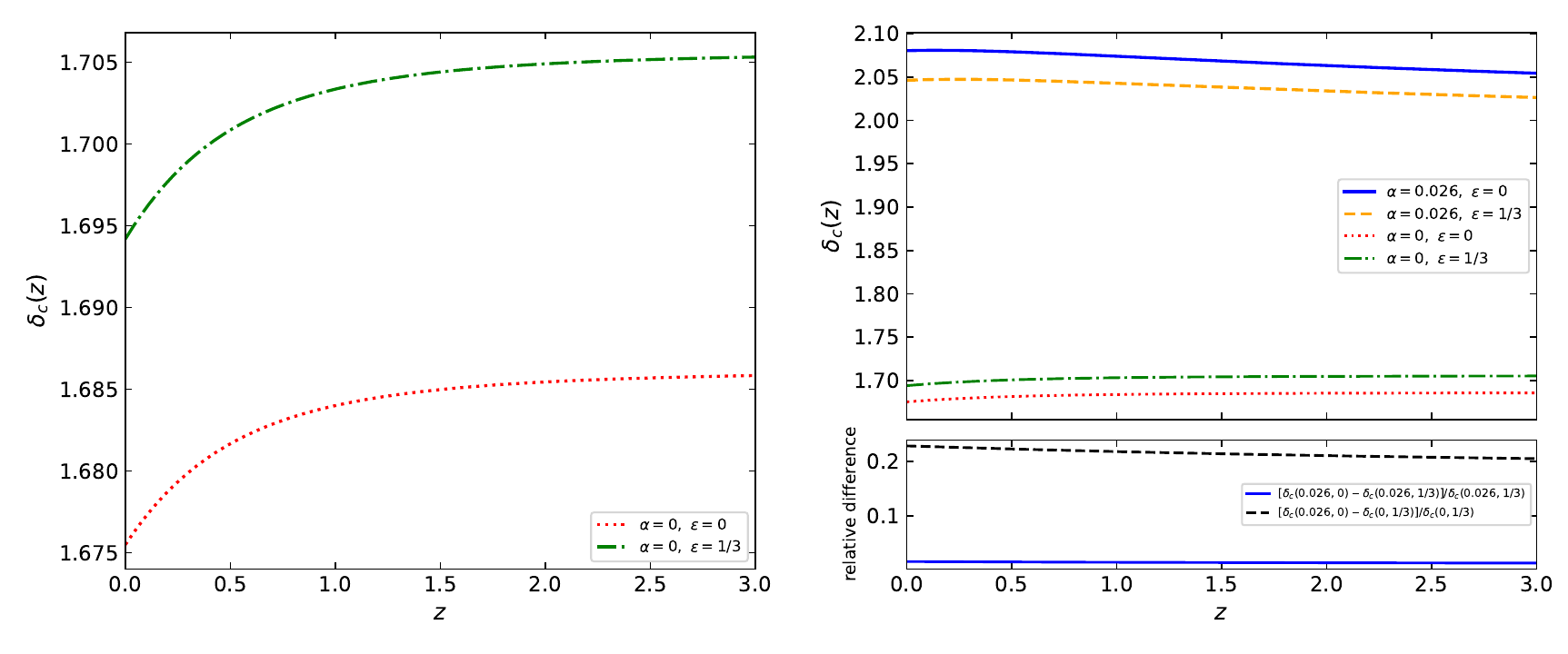}

\caption{Change of $\delta_c$ with redshift. Panel a (left): the upper dot-dashed line represents $\delta_c(z)$ for the LF. The dotted bottom line $\delta_c(z)$ for the SF, GR, limit. Panel b, top solid line SF limit in ESCM, dashed line LF in ESCM. Dot-dashed line LF in SSCM, and the dotted line the SF in SSCM. In the bottom plot are shown the relative differences, as reported.}
\label{fig:deltac_z}

\end{figure*}

The turnaround epoch is defined as the instant at which the radial velocity of the shell
vanishes, $\dot{r}\big|_{a_t}=0$. Using mass conservation, this condition yields the relation
\begin{equation}
\frac{3}{a_t}\left(1+\delta_t\right)=\delta_t',
\label{eq:turnaround}
\end{equation}
where the prime denotes differentiation with respect to the scale factor. The nonlinear density contrast at turnaround is therefore $\delta_t=\delta(a_t)$. At a later stage, the overdensity collapses at the scale factor $a=a_c$. The corresponding
critical overdensity is obtained from the solution of the linearized evolution equation and is
defined as $\delta_c=\delta_{\mathrm{lin}}(a_c)$. Throughout this work, collapse is assumed to occur when the nonlinear density contrast reaches $\delta_{\mathrm{NL}}=\infty$, as in the classical SSCM. We do not follow the choice $\delta_{NL}
=200$ by \citep{Lopes2018} that should be in good agreement with $N$-body simulations and halo mass function calibrations. This choice does not modify the characteristics of the solutions. The pipeline that we follow to find quantities at collapse and turnaround is the following: 1. Choose $M$, $s$, redshift and gravity model ($\epsilon=1/3$, $f_{\rm R0}=10^{-4}, 10^{-5},10^{-6}$). 2. Build the initial Tanh profile. 3. Integrate Eq. \ref{eq:HuSaw} (Eq. \ref{eq:mast}), and the linearized one. 

\begin{figure*}[htbp]
\centering
\includegraphics[width=0.48\textwidth]{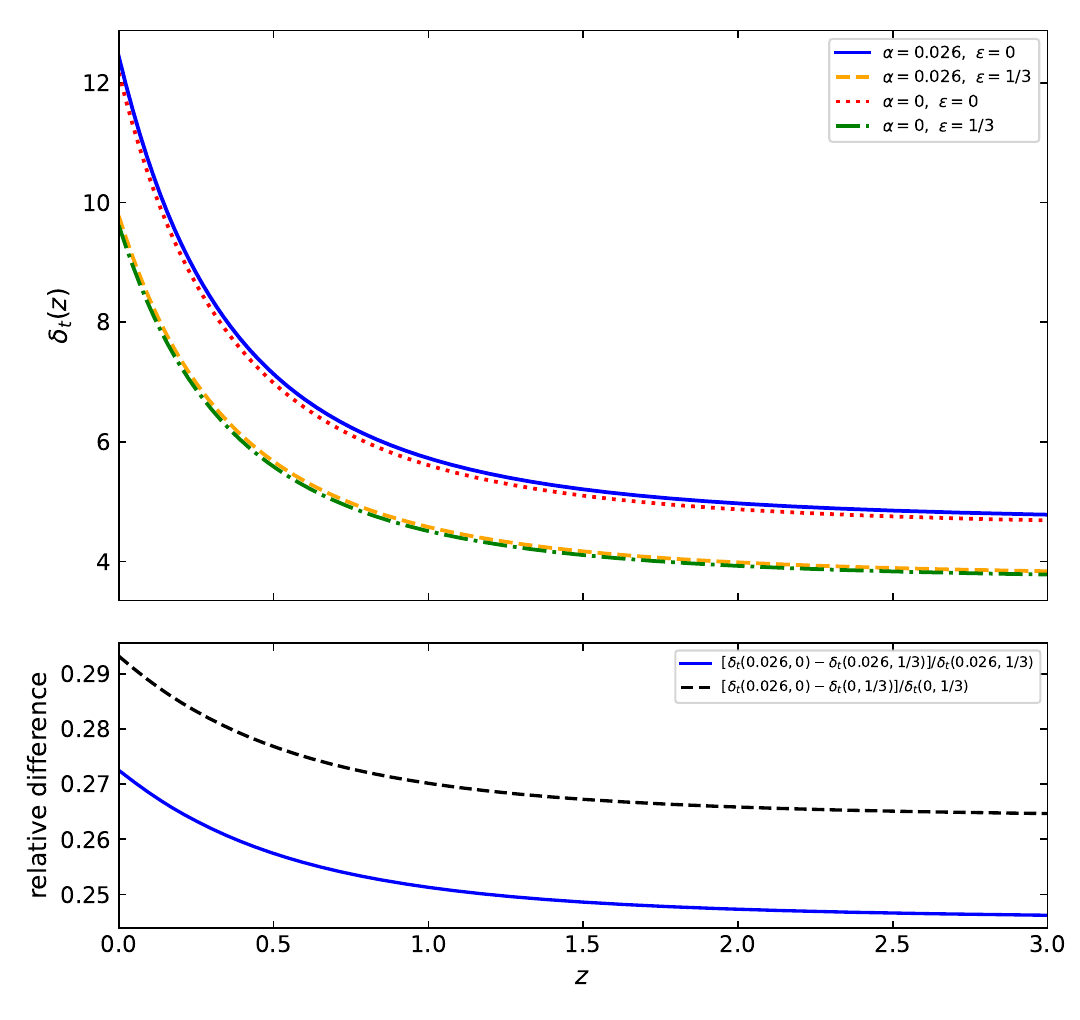}
\caption{Change of $\delta_t$ with redshift. The upper solid line represents the case $\epsilon=0$ (GR) in the ESCM, the dotted line the GR results excluding shear and vorticity. The dashed line the case $\epsilon=1/3$ (LF) in the ESCM. The dot dashed line the result LF excluding shear and vorticity. The bottom panel the relative differences, as reported.}
\label{fig:deltat_z}

\end{figure*}

The nonlinear evolution equation derived in the Appendix C is an integro-differential
equation, since the gravitational potential depends on the Fourier transform of the density
profile. At every integration step the density profile is transformed into Fourier space,
$\delta(r,a) \longrightarrow \delta(k,a)$, where the scale-dependent modification of gravity is evaluated. 
After solving the non-linear equation, the corresponding solution of the linearized equations provides the critical overdensity $\delta_c$. The numerical integration is continued until the turnaround condition $\dot r=0$, is satisfied, which, already written, is equivalent to
\begin{equation}
\delta_t'=\frac{3}{a_t}(1+\delta_t).
\end{equation} 
The corresponding turnaround radius is then obtained from mass conservation.

\begin{figure*}[htbp]
\centering

\includegraphics[width=0.9\textwidth]{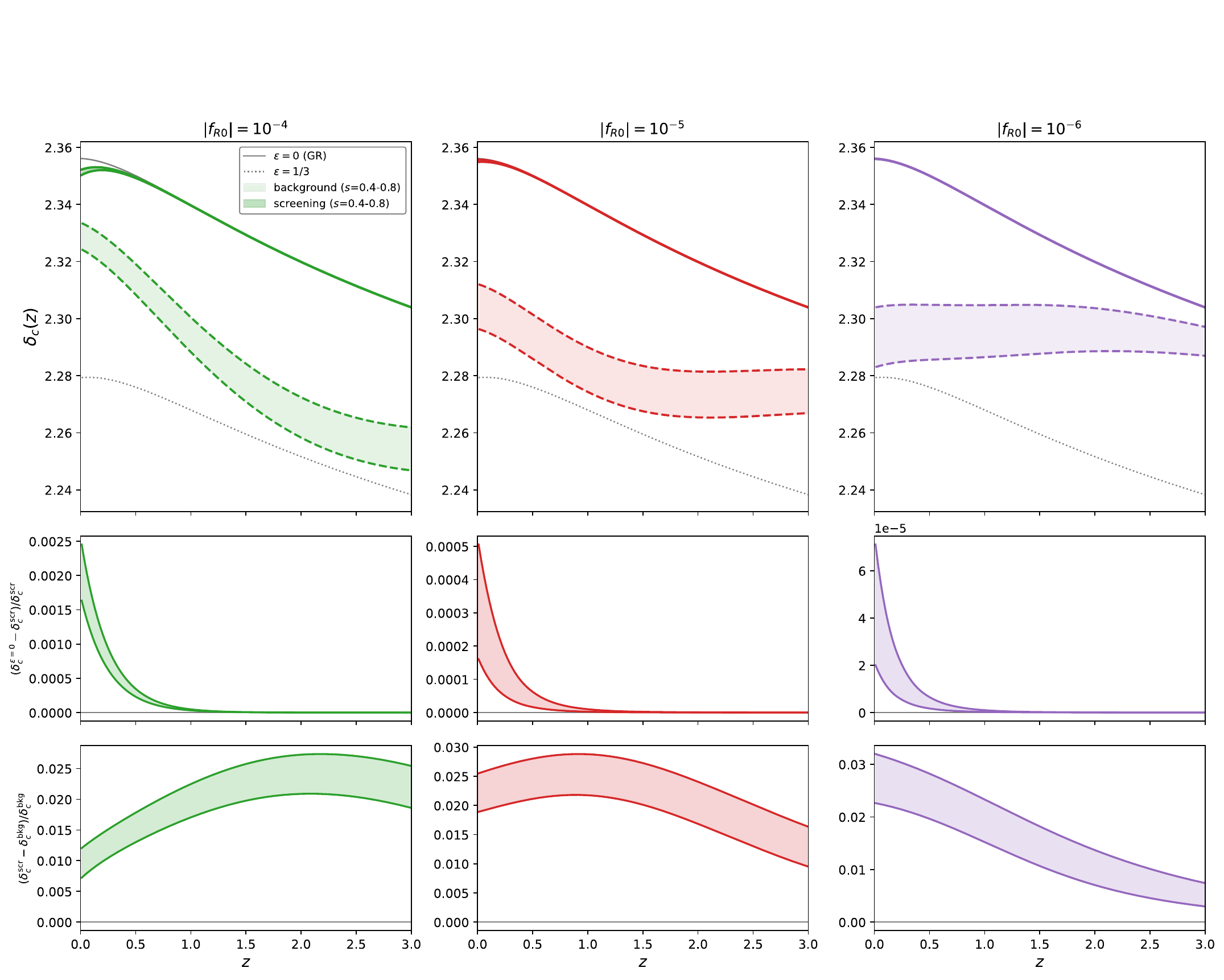}

\caption{The value of $\delta_c$ vs. redshift. The top and bottommost lines represent the screened 
$\delta_c(z)$ for SF and LF in the ESCM, and the shaded solid region $\delta_c(z)$ for F4, F5, and F6, from left to right. The shaded dashed ones, represents the background (no screening). The two bottom rows represent the percentage difference of the quantities reported.}
\label{fig:deltat_screening_z}

\end{figure*}

The density of a spherically symmetric structure with an inner constant mass is given by
$\rho_m(a) = 3M/4\pi R^3(a)$, and the turnaround radius can therefore be computed from the
following relation:

\begin{equation}
R_t(a,M) = \left[\frac{3}{4\pi\Omega_{m0}\rho_c\,[1+\delta_t(a,M)]}\right]^{1/3} a\,M^{1/3}.
\label{eq:Rt}
\end{equation}

Here we used $\rho_m(a) = \bar\rho_m(a)[1+\delta(a,M)]$, with $\bar\rho(a) = \rho_{m0}\,a^{-3}$
and $\rho_c = 2.77 \times 10^{11} h^2\,M_\odot\,\mathrm{Mpc}^{-3}$.

In \citep{DelPopolo2020a}, we showed that in dark energy models the previous equation is generalized, and contains also shear and vorticity 

\begin{equation} \label{eq:turnnp}
R_{\rm ta}= \left [\frac{-3M}{4 \pi \rho_{\rm de}(1+3w)+(\sigma^2-\omega^2)} \right]^{1/3}
\end{equation}
In the case of the $\Lambda$CDM model, ($w=-1$), no shear, and rotation, Eq. (\ref{eq:turnnp}) reduces to
\begin{equation} \label{eq:turn1}
R_{\rm t}= \left [\frac{3GM}{\Lambda} \right]^{1/3}
\end{equation}
which is the same as that obtained by \citep{Pavlidou2014}. 

By definition, $R_t(a,M)$ is the radius enclosing a fixed interior mass $M$ once that
shell stops expanding and starts to collapse, i.e.\ once it reaches turnaround.
Equivalently, $\delta_t(a,M)$ denotes the density contrast evaluated at the precise
instant when the innermost shell of the profile attains its maximum radius. Since this
instant occurs at a specific scale factor, which we call $a_t$, the general quantities
$R(a,M)$ and $\delta(a,M)$ reduce to $R_t(a,M)$ and $\delta_t(a,M)$ exactly when
evaluated at $a=a_t$.

\section{Results}\label{sect:Results}

Several authors have studied the effect of shear \citep{Hoffman1986,Hoffman1989, Zaroubi1993}, and rotation \citep{DelPopolo1998,DelPopolo1999,DelPopolo2000,DelPopolo2001,DelPopolo2002,DelPopolo2013b,Pace2019} on the SSCM and elliptical collapse  model. 

From the relativistic formulation, it follows that vorticity acts as a repulsive force that opposes collapse, while shear accelerates the distortion of the fluid volume, not necessarily preventing the final collapse.

The joint effect of the two quantities changes the turnaround epoch and collapse time, as well as the typical parameters of the SSCM \citep{Peebles1990,Audit1997,DelPopolo2001,DelPopolo2002}. 

As a consequence also the mass function \citep{DelPopolo1999,DelPopolo2000,DelPopolo2013b,Pace2014,Mehrabi2017,DelPopolo2017,Pace2019}, the two-point correlation function, scaling relations like the mass-temperature \citep{DelPopolo2002a,DelPopolo2019}, and luminosity-temperature relation \citep{DelPopolo2005}, are modified.

In this section, we show the results of the calculation, and the impact of the modified gravitational interaction on the
critical density for collapse, $\delta_c$, and on the density contrast at turnaround,
$\delta_t$. We will consider the small-field (SF) regime, where $\epsilon\rightarrow0$ and we are left with the equations in GR and then reobtain GR results.  
The large-field regime, is characterized by $\epsilon\rightarrow1/3$, which represents the maximum enhancement of the gravitational interaction within the linear approximation.

In these two limits, the modified gravitational coupling becomes independent of both scale
and halo mass, allowing a direct comparison with the standard $\Lambda$CDM case. The
quantities $\delta_c$ and $\delta_t$ are evaluated for structures collapsing at the present
epoch ($z=0$, $a=1$) using the Tanh  initial density profile.

To quantify the effect of modified gravity, we compute the fractional variation of each quantity with respect to the
GR prediction. The lines in Fig. 1a (by a and b we mean the left and right panels), 
represent the values of $\delta_c$ in the redshift range $0<z<3$
and for $10^{14} h^{-1} M_{\odot}$. The curves correspond to those in Fig. 1 of \citep{Lopes2018}, 
with the difference that they used as collapse $\delta_{NL}=200$, and we used a larger value. 

These curves are those typical of spherical collapse.
The two lines of Fig. 1a show that the dot-dashed curve corresponding to a larger gravity ($\epsilon=1/3$) and dotted one which is the $\Lambda$CDM counterpart, 
are characterized by the fact that the collapse in presence of larger gravity gives a larger $\delta_c$. 

This might seem strange since a larger gravity should give rise to a faster collapse and a smaller $\delta_c$. In reality, the critical threshold $\delta _{c}$ does not represent the non-linear density of the collapsing structure. Instead, it is the density value that the perturbation would have reached if it had evolved purely according to linear growth theory, evaluated precisely at the moment the real, non-linear system completely collapses. When gravity is stronger (for instance, due to a fifth force or a modified gravity parameter like $\epsilon =1/3$), non-linear collapse is faster and the linear growth is even faster. Since the stronger gravity accelerates the linear growth during the early stages even more than it hastens the final non-linear collapse, the fictitiously extrapolated linear density $\delta$ reaches a higher value by the time the actual collapse happens. In summary: the curve for stronger gravity is higher because the enhanced gravitational force drives linear perturbations to grow at an accelerated rate. By the time the real object forms, the linear theory has already extrapolated the density to a higher threshold than it would have reached under standard gravity. 

The top solid and dashed line, in Fig. 1b, are obtained by means of ESCM, while the bottom dot-dashed and dotted lines in the SSCM ($\alpha=\epsilon=0$), are the same as in Fig. 1a. The collapse threshold $\delta_c$ for $\epsilon=0$ in the ESCM, differs from $\delta_c$ with $\epsilon=1/3$ by around $1.7 \%$, 
but the difference is larger than $20 \%$ 
 if we consider the difference between the ESCM result for $\epsilon=0$ (GR), and the $\epsilon=1/3$ in the SSCM (dot-dashed line).  

The two top lines illustrate how the critical linear overdensity for collapse, $\delta_c(z)$, is affected by the combined action of shear and vorticity (parameterized by $\alpha$) and by the modification of gravity (parameterized by $\epsilon$). The large separation between the curves with
$\alpha=0.026$ (the value of $\alpha(M)$, related to shear and vorticity, at $10^{14} h^{-1} M_{\odot}$) (solid and dashed) 
and those with $\alpha=0$ (dot dashed and dotted), reflects the stabilizing action of shear and vorticity. Non-radial motions generate an effective support against gravitational collapse, reducing the efficiency of the inward motion of matter. Consequently, a perturbation must start from a larger initial linear overdensity in order to collapse at the same redshift. 
The position of the upper curves is determined by the competition between the stabilizing effects of shear and vorticity and the enhancement of gravity produced by the modified-gravity parameter $\epsilon$. Shear and angular momentum oppose the radial infall of matter, increasing the critical linear overdensity required for collapse. As a result, the models with 
$\alpha=0.026$ are systematically shifted toward higher values of $\delta_c$ than the corresponding spherical-collapse models with $\alpha=0$. For a fixed value of $\alpha$, the case $\epsilon=1/3$ strengthens the effective gravitational interaction. 

While among the two bottom lines the one with $\epsilon=1/3$ lies above that with $\epsilon=0$, in the two top curves the effect of shear and vorticity is larger than that of enhanced gravity, so the curve ($\alpha=0.026$, $\epsilon=0$) lies above the one corresponding to larger gravity ($\epsilon=1/3$).

The redshift evolution is also physically meaningful. Shear and vorticity, as already shown in \citep{DelPopolo2013a,DelPopolo2013b,Pace2014,Mehrabi2017,Pace2019,DelPopolo2020a,DelPopolo2020b}, modify the behaviour of $\delta_c$, which becomes redshift dependent. The curves with $\alpha=0.026$ decrease slowly with increasing redshift, indicating that the additional support provided by shear and angular momentum becomes progressively less effective as the Universe approaches the matter-dominated regime. In contrast, the two curves with $\alpha=0$ remain almost constant, showing that the standard spherical-collapse threshold depends only weakly on redshift. Overall, the upper panel demonstrates that the dominant effect is produced by shear and vorticity, which increase the collapse threshold, whereas the modification of gravity introduces only a relatively small correction.

In the bottom panel is shown the percentage difference between the top curves, namely the $\Lambda$CDM $\delta_c$ (solid) line
and the threshold affected by MG, both obtained with the ESCM. The difference, as already discussed, is around $1.7 \%$, slightly larger than \citep{Lopes2018} result. The difference between the $\Lambda$CDM in ESCM (solid line), and that of the SSCM with $\epsilon=1/3$ is much larger, $\simeq 20\%$. 

\begin{figure*}[htbp]
\centering

\includegraphics[width=0.9\textwidth]{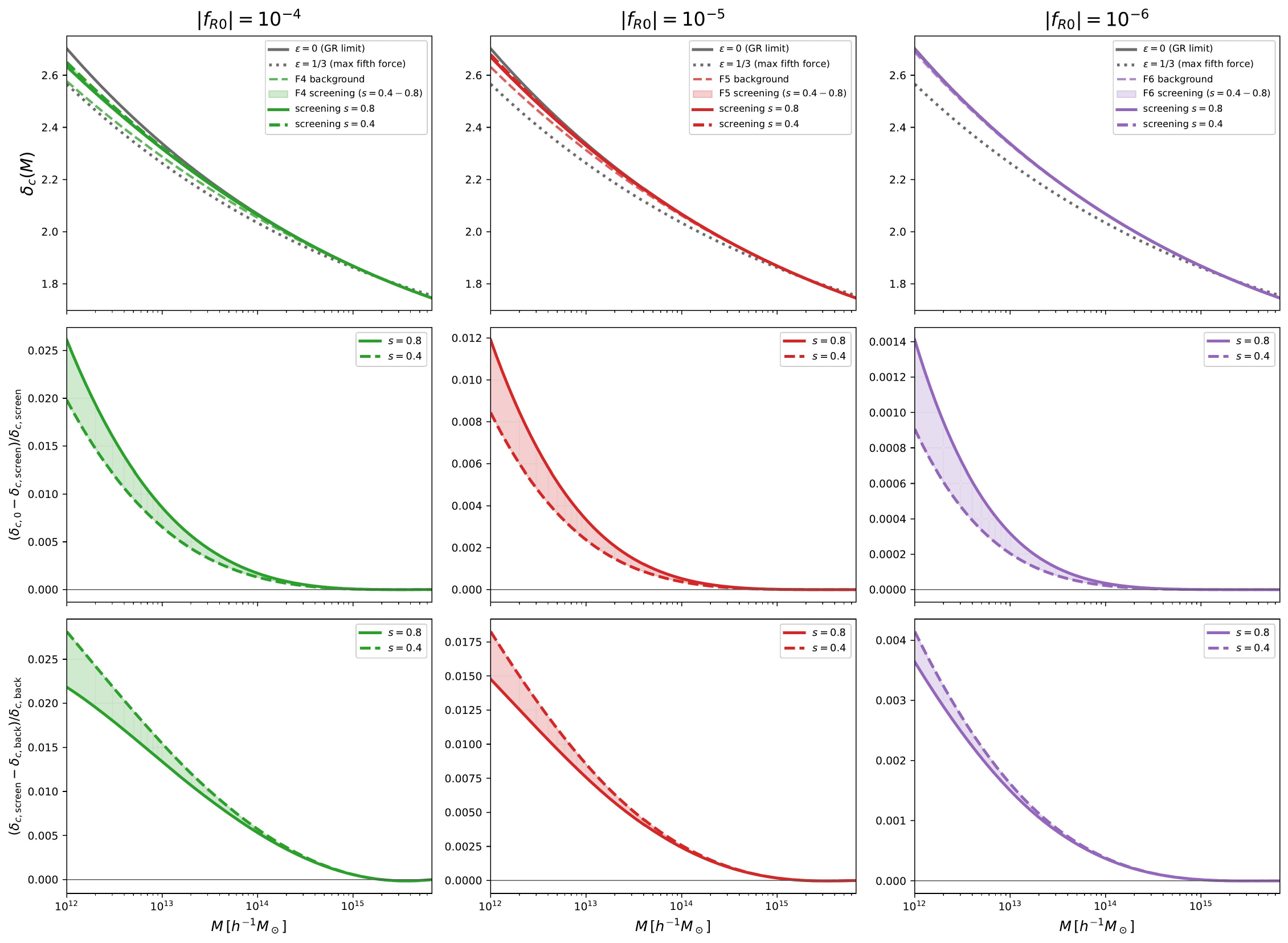}

\caption{Same as Fig. 3 but for $\delta_c(M)$.}
\label{fig:deltac_M}

\end{figure*}

Fig. 2 plots $\delta_t$ in terms of $z$ in the same redshift range and the same mass as in Fig. 1.

In this case the value of $\delta_t$ for $\Lambda$CDM model (dotted line), is larger than that for $\epsilon=1/3$ (dot-dashed line). This is due to the fact  that turnaround happens before, when gravitation is larger with respect to the $\Lambda$CDM case, where gravitation is smaller. Non-linear collapse happens more easily, since gravity is stronger, smaller initial seed perturbations are required to form a structure. In the ESCM model, the solid line, GR, and the dashed line, $\epsilon=1/3$ case, is similar to Fig. 1. The difference between these two curves is around $27 \%$ slightly larger than \citep{Lopes2018}, while the difference between the solid and dot-dashed line is $\simeq 29 \%$. As in Fig. 1, the curves with $\alpha=0.026$ sit above the curves with $\alpha=0$. A higher threshold means that collapse is more difficult, since the parameter $\alpha$ (shear and vorticity) acts as an opposing effect. It requires a denser initial patch to overcome this resistance and successfully collapse. 

Moving to the right (higher $z$), going deeper into the past, the turnaround overdensity $\delta_t(z)$ decreases
"sharply". 
{In the early universe (higher $z$), the background universe is much denser and entirely matter-dominated, making it much easier for perturbations to grow and collapse. Therefore, a lower linear density threshold is sufficient to trigger collapse. As the universe expands toward the present day ($z \rightarrow 0$), dark energy or cosmic acceleration begins to dominate ($z \simeq 0.3-0.4$), heavily resisting structure formation. This requires a much higher overdensity threshold to achieve collapse at later times.
}

\begin{figure*}[htbp]
\centering

\includegraphics[width=0.9\textwidth]{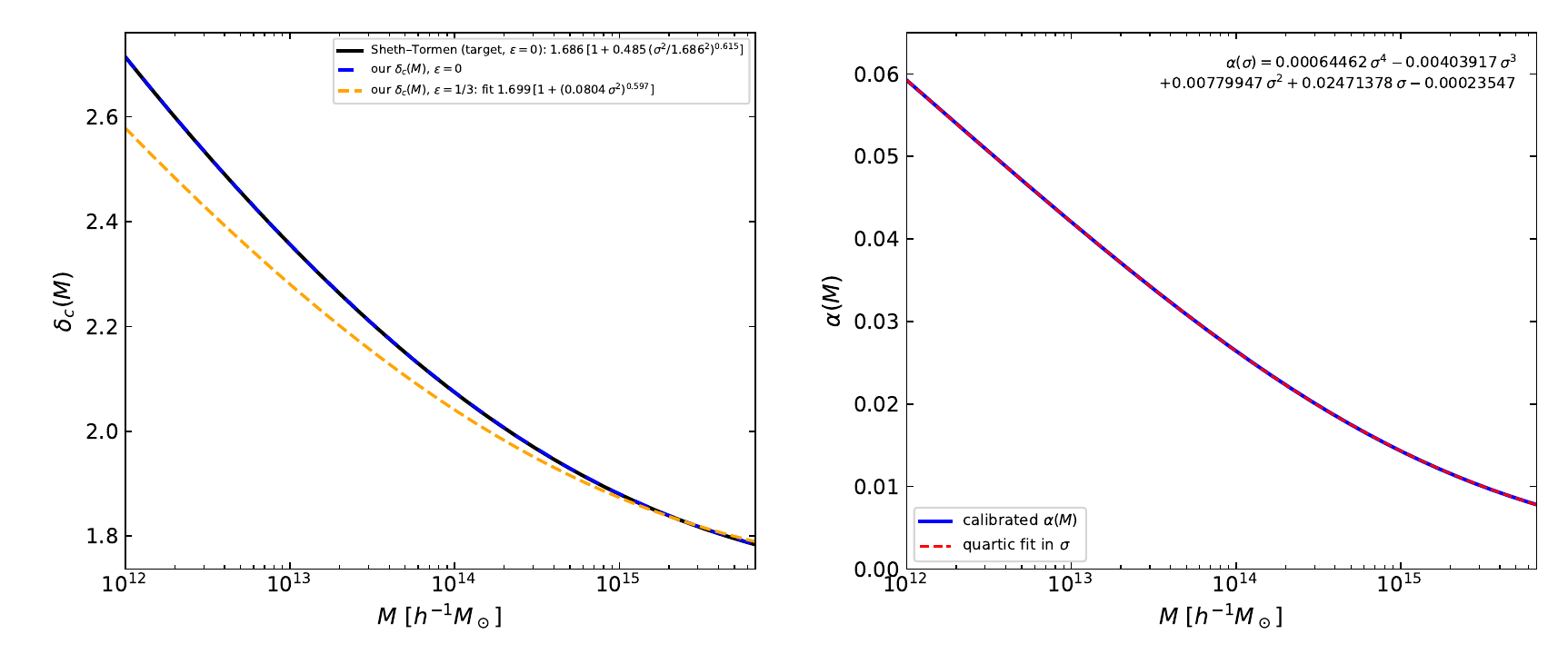}
\caption{Left panel: Trend of $\delta_c(M)$ for GR in ESCM (dashed blue line), comparison with \citep{SMT2001}, solid line, and $\delta_c(M)$ for the LF in ESCM, dashed orange line. Right panel: the mass dependence of $\alpha$, where $\sigma(M)$ is the rms fluctuation. The solid line is the calibrated $\alpha(M)$, and the dashed line, the quartic fit.}
\label{fig:deltac_alpha_M}

\end{figure*}

In the following plots, we do not consider only the two limiting cases $\epsilon=1/3$ (LF), and $\epsilon=0$ (SF), but we also consider the changes of MG at different scales, and different times, through $\epsilon(k,a)$. This implies an influence of one scale on the others. Eq. \ref{eq:HuSaw} (\ref{eq:mast})  becomes an integro-differential equation 
and one has to solve the Fourier integral at each step in time as described in Appendix C. Due to the existence of a further scalar degree of freedom in $f(R)$, allowing monopole radiation, Birkhoff's theorem is no longer valid. Then, the dependence of $\delta_c$ and $\delta_t$ from $|f_{R0}|$ will bring to a mass dependence of those quantities. As already reported, we consider $|f_{R0}|=10^{-4}, 10^{-5}, 10^{-6}$, and mass in the range $10^{12} h^{-1} M_{\odot}-6.7 \times 10^{15} h^{-1} M_{\odot}$ 

In this mass range one has to expect effects due to the chameleon mechanism. Masses smaller than $10^{11} M_{\odot}$, characterized by shallow gravitational potential wells, 
live in the unscreened regime. Chameleon mechanism does not activate, modified gravity is fully active, making collapse easier. For large masses, $>10^{16} M_{\odot}$ potential wells are extremely deep (massive galaxy clusters), and consequently 
Chameleon is active, and we are in the screened regime. The fifth force is suppressed, and the system reverts entirely to standard GR ($\epsilon = 0$). 

The model proposed by \citep{Lopes2018}, and extended in this paper, computes the turnaround radius and the collapse threshold within Hu--Sawicki $f(R)$ gravity by solving the nonlinear evolution equation for the density contrast, Eq. \ref{eq:HuSaw} (
Eq. \ref{eq:mast})(with $\alpha=0$). The effects due to the chameleon mechanism are not considered 
by \citep{Lopes2018} since according to them the overdensities at the turnaround radius (their goal is to study this phase to disentangle between GR and MG)  are typically of order one. 
Chameleon screening switches on when the local density is much higher than the cosmic mean. If the overdensity at the turnaround shell is only ``of order one'' (the classical, non-linear spherical-collapse value for an EdS universe is $\delta \approx 9\pi^2 / 16 - 1 \approx 4.55$, in contrast to virialization, where $\delta \sim 100$ or more), then there is not yet a deep enough potential well to efficiently activate screening -- this is a common simplification in that literature (turnaround radius as a cosmological probe). In reality, $\delta_t$  is in the range $9-12.5$, and with shear and vorticity (ESCM) the value is even higher than in the classical case. As we will see in Fig. 7, the correction due to screening goes from values $>20 \%$ (F4) to $10 \%$ (F5), thus screening is not negligible. In Appendix D, we discuss how we take the mechanism into account. 

Fig. 3, plots the collapse threshold, $\delta_c$, in terms of the redshift, $z$. The top three plots of the figure show how the threshold varies with $z$ for  $f_{\rm R0}=10^{-4}, 10^{-5}, 10^{-6}$ (hereafter F4, F5, F6), going from left to right. The
solid line represents GR (SF), the dashed bottom one the LF limit. The regions between dashes represent the background (no screening correction). The initial profile we used is Tanh and we considered two values for $s$, $s=0.4$ and $s=0.8$, between which is shown a shaded region. 
The solid curves represent $\delta_c(z)$ taking into account screening. Going from left to right, namely from $F4$ to $F6$, the screening pushes the background curves towards the GR limit, till $f(R)$ coincides with GR. The bottom row shows the percentage difference between GR and screened $\delta_c$, which at F4 is $0.25 \%$ and tends to zero going to the right (F6). The third row, represents the percentage difference between screened $\delta_c$ and the background. The difference increases toward the right. The percentage difference between $\epsilon=0$ and the background at F4 is a bit larger than $1 \%$, and $3 \%$ at F6.

Fig. 4 shows the mass dependence of $\delta_c$. Solid and dotted lines represent the SF and LF limits. The shaded regions the screened $\delta_c$ (F4, F5, and F6, from left to right). The quoted regions tend towards GR, going from F4 to F6. The dashed lines represent the background, while the bottom curves represent the percentage difference between GR and screened $\delta_c$, showing how screened $\delta_c$ moves to the GR values. The maximum difference between the GR limit and the screened $\delta_c$, at $10^{12} M_{\odot}$, for F4 is $ 2.5 \%$. The last row represents the percentage difference between the screened $\delta_c$ and the background. Both change with $f_{\rm R0}$, and their difference decreases from F4 to F6.
It is important to note that the mass dependence comes both from shear-vorticity, and the MG. In the other papers \citep{Kopp2013,Borisov2012,Li2012,Lam2012,Cataneo2016,Lopes2018}, excluding simulations \citep{HernandezAguayo2018,Gupta2022}, the effect of shear and vorticity is neglected. Of the quoted papers, \citep{Kopp2013} is the most rigorous: it performs numerical simulations of spherical collapse by solving the quasi-static $f(R)$ equations, including the non-linear scalar field. Then, they build a fit to their spherical collapse results, to get $\delta_c(M,z,f_{R0})$, however their system is isolated, it does not consider interaction with neighbors, and then neglects shear and vorticity.  

\begin{figure*}[htbp]
\centering

\includegraphics[width=0.9\textwidth]{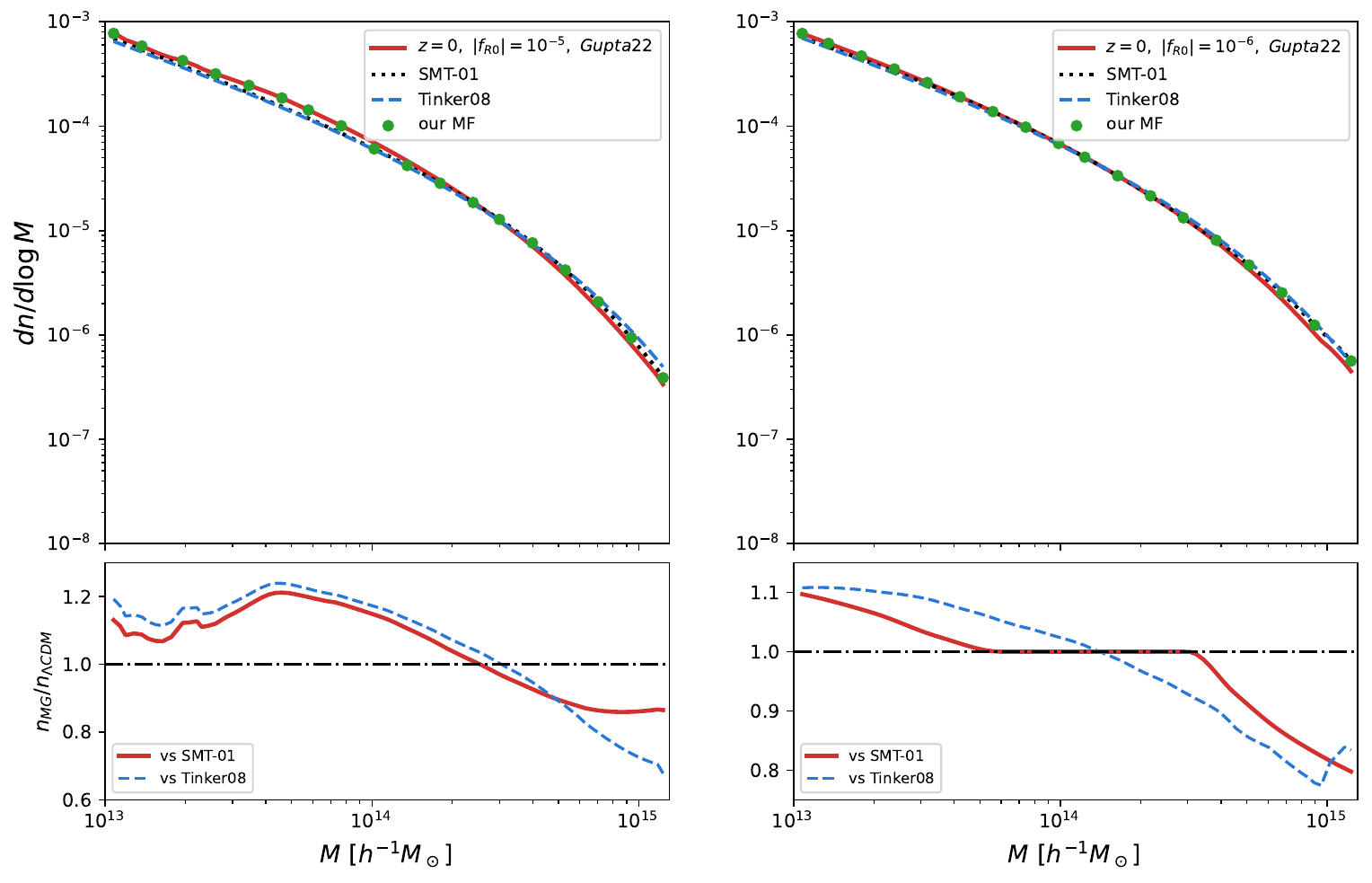}

\caption{Mass function for F5, and F6. The solid line corresponds to \citep{Gupta2022,Gupta2022a}. The dashed and dotted lines  
\citep{Tinker2008}, and \citep{SMT2001} mass function. The dots represent our mass function. }

\label{fig:massfunction}

\end{figure*}

Then, in those papers the mass dependence of $\delta_c$ comes just from the MG. In reality, mass dependence must appear in the collapse even in GR as shown by \citep{DelPopolo1998a,SMT2001,DelPopolo2007}. The collapse threshold is a decreasing function of mass as seen in Fig. 4. We show this behavior more clearly in Fig. 5a. The dashed blue line represents $\delta_c(M)$ in our model, the solid black line the \citep{SMT2001} threshold. This kind of dependence was shown for the first time by \citep{DelPopolo1998}, considering that structures are subject to tidal torque effects, and  
by \citep{SMT2001}, using an ellipsoidal collapse, and extended by \citep{DelPopolo2017} to take into account other effects. 
Since the threshold is fundamentally, apart from some numerical normalization factors, the barrier used in the Excursion Set Theory to calculate the Dark Matter Halo Mass Function, its correct calculation is important. If we look at papers like \citep{Kopp2013,Borisov2012,Li2012,Lam2012,Cataneo2016}, shear and vorticity are not taken into account and the collapse threshold is an increasing function of mass to reach the classical GR value. In those papers, the mass dependence in the barrier comes only, as already reported, from MG, since the interaction of a protostructure with neighboring ones 
it is not taken into account, or the elliptical collapse extension. Consequently, a part of the mass dependence is lost, in those approaches, and this influences the mass function. The effect of shear and vorticity has been studied in \citep{DelPopolo2013,DelPopolo2013a,DelPopolo2013b,Pace2014,Mehrabi2017}. In our model shear and vorticity are parameterized by a mass dependent term, $\alpha(M)$, see Eq. \ref{eq:sigm} and \citep{DelPopolo2013,DelPopolo2013a,DelPopolo2013b,Pace2014,Mehrabi2017}. It is a monotonically decreasing function of mass, roughly estimated at galactic and cluster scales and having values of 0.05 \citep{DelPopolo2013} at MW scale,  
and 0.005 at clusters scale \citep{Mehrabi2017} (in Appendix B, there is a short discussion on $\alpha$ values). 
In \citep{Pace2018}, there is an attempt to obtain an estimate of $\sigma^2-\omega^2$, but as they report, some limitations in their approach make a direct comparison with \citep{DelPopolo2013} difficult. So, as in \citep{DelPopolo2013} and subsequent papers we chose to obtain the complete mass dependence of $\alpha$, assuming that the collapse threshold of our model fits that of \citep{DelPopolo1998,SMT2001}. Our $\alpha(M)$ is plotted in Fig. 5b. 

In order to reobtain Kopp's threshold, it is necessary to set $\alpha=0$, namely discard shear and vorticity.

Once the threshold, which represents the barrier, is known, $\delta_c=B(S)$ (where $S$ is the variance of the linear density field smoothed on a mass scale $M$)\footnote{In \citep{Sheth2002}, instead of simply considering $\delta_c=B(S)$, the barrier was slightly modified by multiplying the threshold of spherical collapse, $\delta_{c,sc}=1.686$, by $\sqrt{a}$: $B(S) = \sqrt{a}\,\delta_c \left[ 1 + b \left( \frac{S}{a\,\delta_c^2} \right)^c \right]$, to improve the agreement with simulations}, it is possible to obtain the mass function. 
In excursion set theory, mapping the formation of cosmic structures reduces to a first-crossing time problem. 
For a non-flat, moving barrier such as the one obtained, an exact analytical solution for the first-crossing distribution does not exist. To overcome this limitation, the problem may be formulated via a Volterra integral equation or utilizing numerical Monte Carlo (MC) simulations of random walks. Using MC simulations, one can obtain the multiplicity function, $\nu f(\nu)$, the fraction of total cosmic mass that collapses into halos per unit interval of the scaling variable, where $\nu \equiv \frac{\delta_c}{\sigma}$. Finally the mass function is given by:
\begin{equation}
n(M) dM = \frac{\bar{\rho}}{M^2} \left[ \nu f(\nu) \right] \left\vert \frac{d \ln \nu}{d \ln M} \right\vert dM
\end{equation}

\begin{figure*}[htbp]
\centering

\includegraphics[width=0.9\textwidth]{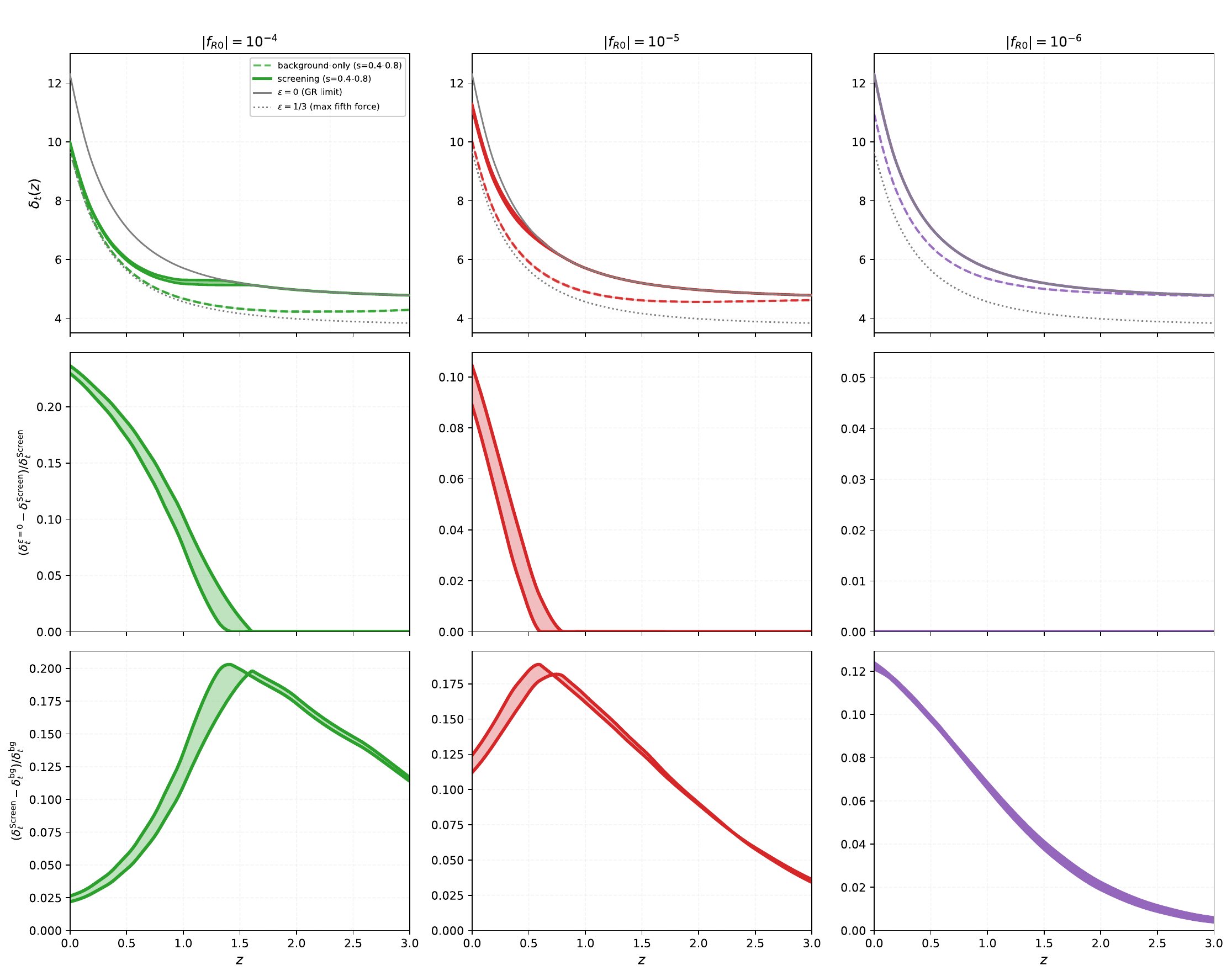}

\caption{Changes of $\delta_t$ with $z$. Symbols as in Fig. 4. }
\label{fig:deltat_screening_z2}

\end{figure*}

Using the thresholds plotted in Fig. 4 for F4, F5, and F6, 
obtained using the ESCM, we obtain the mass function by means of MC simulations. The result is plotted in Fig. 6. We consider the cases F5 and F6 plotted in Fig. 2 of Gupta \citep{Gupta2022a} (see also \citep{Gupta2022}), for $z=0$\footnote{F4, as already reported is excluded from observational constraints.}
Gupta's MG mass functions were obtained using the dark matter halo catalogs generated using Elephant (Extended LEnsing
PHysics using ANalaytic ray Tracing). The dashed line represents Tinker's mass function \citep{Tinker2008}, the dotted line the mass function given in \citep{SMT2001}, and the green dots the mass function that we obtained. The right panel is the case F6. We considered only these two cases because of the constraints on the values of $f_{\rm R0}$:  $f_{\rm R0}<10^{-4.79}$
\citep{Cataneo2015}, and $f_{\rm R0}<10^{-4.31}$ \citep{Artis2024}.

In the Hu-Sawicki $f(R)$ modified gravity model, the background field values $|f_{R0}| = 10^{-5}$ and $10^{-6}$ lead to an enhanced abundance of massive galaxy clusters compared to  GR/$\Lambda$CDM, leading to an increase of the 
mass function. 
The magnitude of this deviation and the mass scale at which it occurs depend heavily on the competing effects of enhanced growth and the chameleon screening mechanism, and is more visible at high redshift (see \citep{Gupta2022,Gupta2022a}, and Fig. 1 of \citep{HernandezAguayo2018}, for $z=0.5$). As also shown in \citep{Gupta2022a} for higher redshifts, for the case F5, F6, the $f(R)$ mass function predicts more and more haloes. 
 
In low-density regions, such as cosmic voids and large-scale structures, the scalar degree of freedom (the ``fifth force'') is unscreened. The fifth force increases the growth rate of linear perturbations with the consequence of boosting the overall normalization of density fluctuations, effectively increasing $\sigma_8$ relative to $\Lambda$CDM. Consequently, a higher number of massive halos forms per unit volume compared to GR.
In high-density environments, (small scales, and halo centers) the scalar field couples to a larger effective mass ($m_{\rm eff}$), suppressing the fifth force via the chameleon mechanism, and the effective gravitational constant returns to its Newtonian value ($G_{eff} \approx G_{N}$).

For $|f_{R0}| = 10^{-5}$, the Compton wavelength of the scalar field is relatively large, and the field is lighter. The chameleon screening is less efficient, failing to screen galaxy clusters and intermediate-mass halos. This results in a  widespread enhancement in the halo mass function (HMF) across a broad range of masses ($\gtrsim 10^{14} M_\odot$).
For $|f_{R0}| = 10^{-6}$, the fifth force has a shorter range. The chameleon screening mechanism becomes highly efficient and screens most typical galaxy clusters. As a result, the HMF closely tracks the standard GR predictions, with detectable enhancements confined only to the most massive, rarest clusters that remain partially unscreened.

In a future paper, we intend to study more in detail $\delta_c(M,z,f_{R0})$, and the mass function, using Kopp's approach, and adding the effect of shear and vorticity.   

We now proceed with the discussion of the characteristic quantities in the SSCM and in our case in the ESCM. In Fig. 7, we plot the evolution with redshift of the turnaround overdensity. The top panels refer to $\delta_t(z)$ for the SF and LF limit, the background, dashed lines, and the effect of screening on $\delta_t(z)$, shaded region, and this for F4 to F6, from left to right. As one can see, the screening changes $\delta_t(z)$ behaviour, and the last quantity tends to be pushed towards the SF (GR) limit, going from F4 to F6. The bottom row represents the percentage difference between the value of $\delta_t(z)$ in the SF limit and that of screening. In the last row, the percentage difference between screening and background. In some papers as \citep{Lopes2018}, the large difference between the SF limit (GR) and the background ($\simeq 30 \%$), and larger than $20 \%$  for the difference between the SF limit (GR) and the screened value, has led to think that turnaround and related quantities could be used to distinguish between GR and MG. As Fig. 7 shows, this difference decreases till $0 \%$ going toward F6, and so these quantities could be used to distinguish only between some values of $f_{\rm R0}$: F6 is excluded because too close to GR, and F4 as we have already seen  is excluded by several constraints already cited. So, only the F5 case remains. In summary, stating that $\delta_t$ can be used to distinguish between GR and MG is not exactly true (in the following we will see another constraint coming from simulations \citep{Hansen2020}). Then we may go directly to turnaround, showing its dependence on $z$ and mass.  

\begin{figure*}
\includegraphics[width=0.8\textwidth]{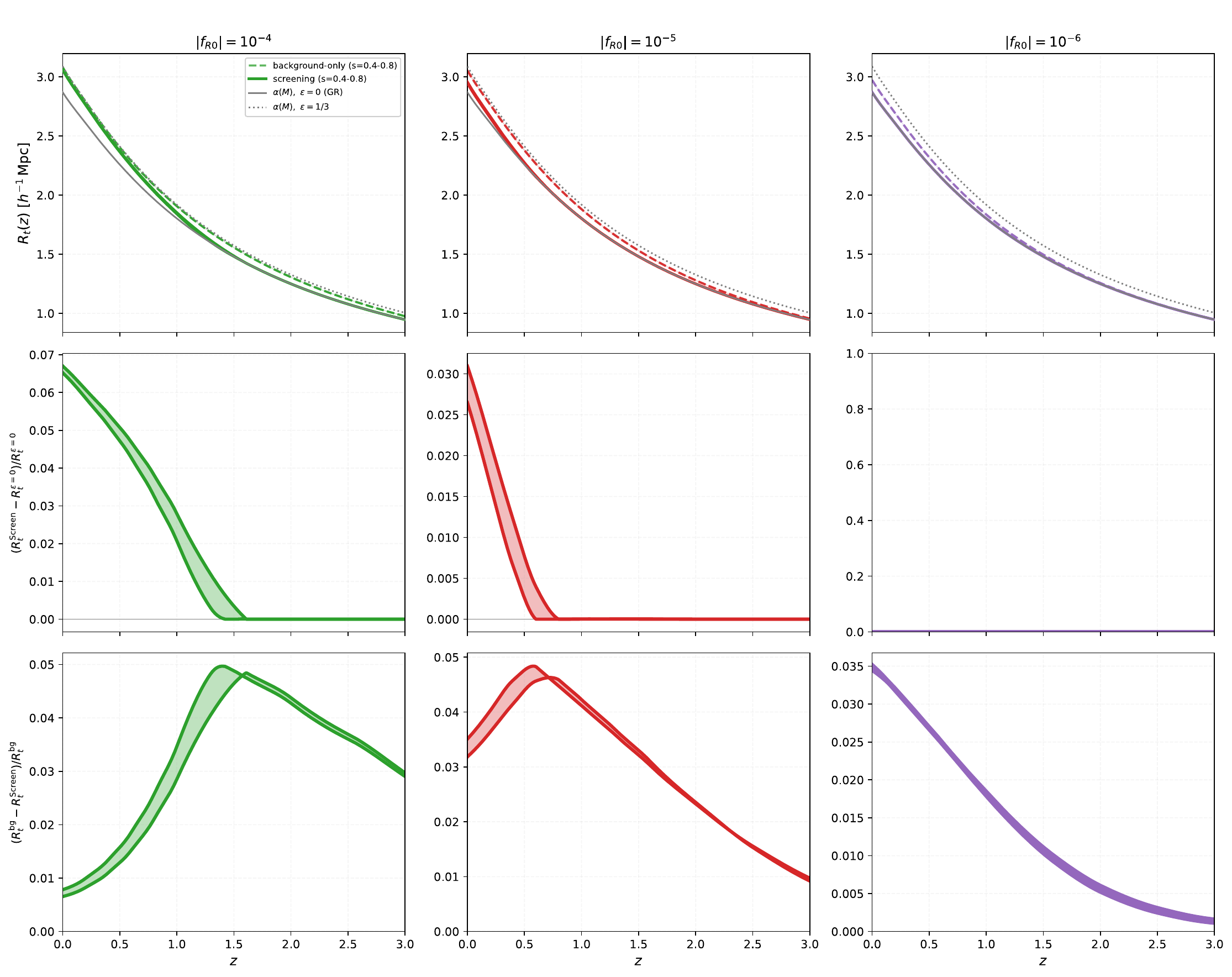}
\hfill
\includegraphics[width=0.75\textwidth]{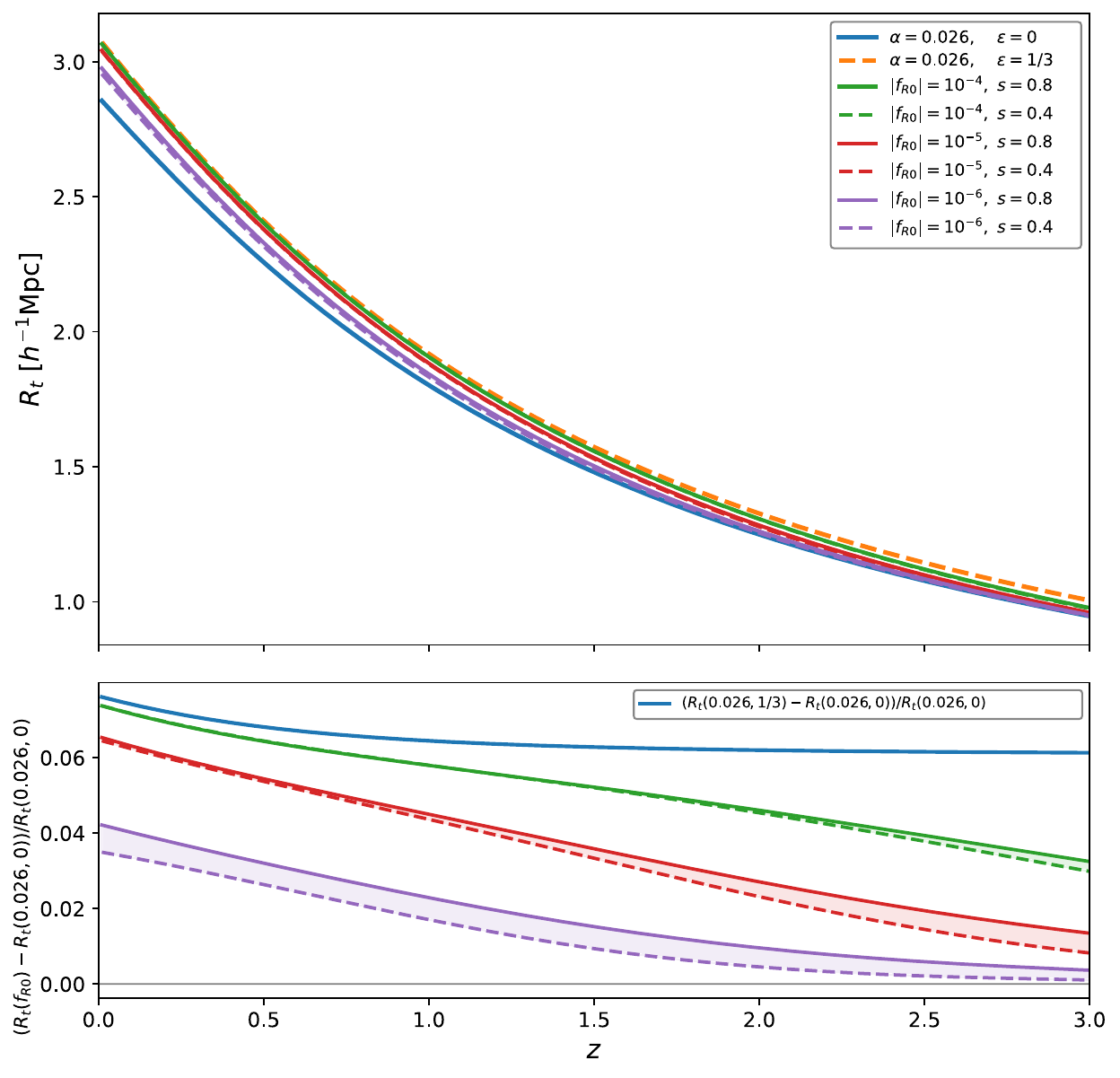}

\caption{Top panel: Variation of $R_t$ with redshift. Symbols as in Fig. 4. Bottom panel: same as the previous panel, but showing only the background values of F4, F5, and F6.}
\label{fig:Rt_z}

\end{figure*}

In Fig. 8, we plot the dependence of turnaround on redshift, $R_t(z)$. Fig. 8, top plot, is similar to Fig. 7. The top panels show the evolution of turnaround for LF, SF, and $f(R)$. Both the background, and the effect of screening for F4, F5, and F6, are shown. The bottom rows are similar to those in Fig. 7. 

When we go to higher $z$, we are looking further back in time, when the ambient cosmic matter density was 
significantly higher. This rapid density increase at high $z$ drives the decay features across all rows.

In $f(R)$ modified gravity theories, the effective mass of the scalar field degree of freedom (the scalaron) depends heavily on the local curvature and density. At higher redshifts, the increased density $\rho(z)$ dramatically increases $m_{\text{eff}}$. This severely limits the range of the modified gravity fifth force ($\lambda_C \propto m_{\text{eff}}^{-1}$). Consequently, the characteristic transition scale $R_t(z)$ must steadily decay and shrink as $z \to 3$.

The fractional differences drop sharply to zero at a threshold redshift ($z \approx 0.7$ for $|f_{R0}| = 10^{-5}$ and $z \approx 1.5$ for $|f_{R0}| = 10^{-4}$). At high $z$ (Early Universe) the background density exceeds the threshold required to activate screening. The fifth force is entirely quenched, forcing the modified model to become identical to standard GR. Thus, $\Delta R_t \to 0$.
At low $z$ (Late Universe), when the universe expands and dilutes, the density drops below the critical threshold, allowing the screening to break down and letting modified gravity deviations grow.

In the bottom row, the relative deviations against the smooth background model peak at a specific cosmic epoch and then decay at higher redshifts. As screening takes full effect at early times, the local perturbations are suppressed. The system smoothly converges back toward the unperturbed background metric, forcing the relative ratio toward zero at the right side of the plots. The decay rate and the activation redshift of the screening depend on the background field value $|f_{R0}|$:
\begin{equation}
|f_R(z)| \approx |f_{R0}| \left[ \frac{\bar{R}_0}{\bar{R}(z)} \right]^2
\end{equation}

\begin{figure*}
\includegraphics[width=0.9\textwidth]{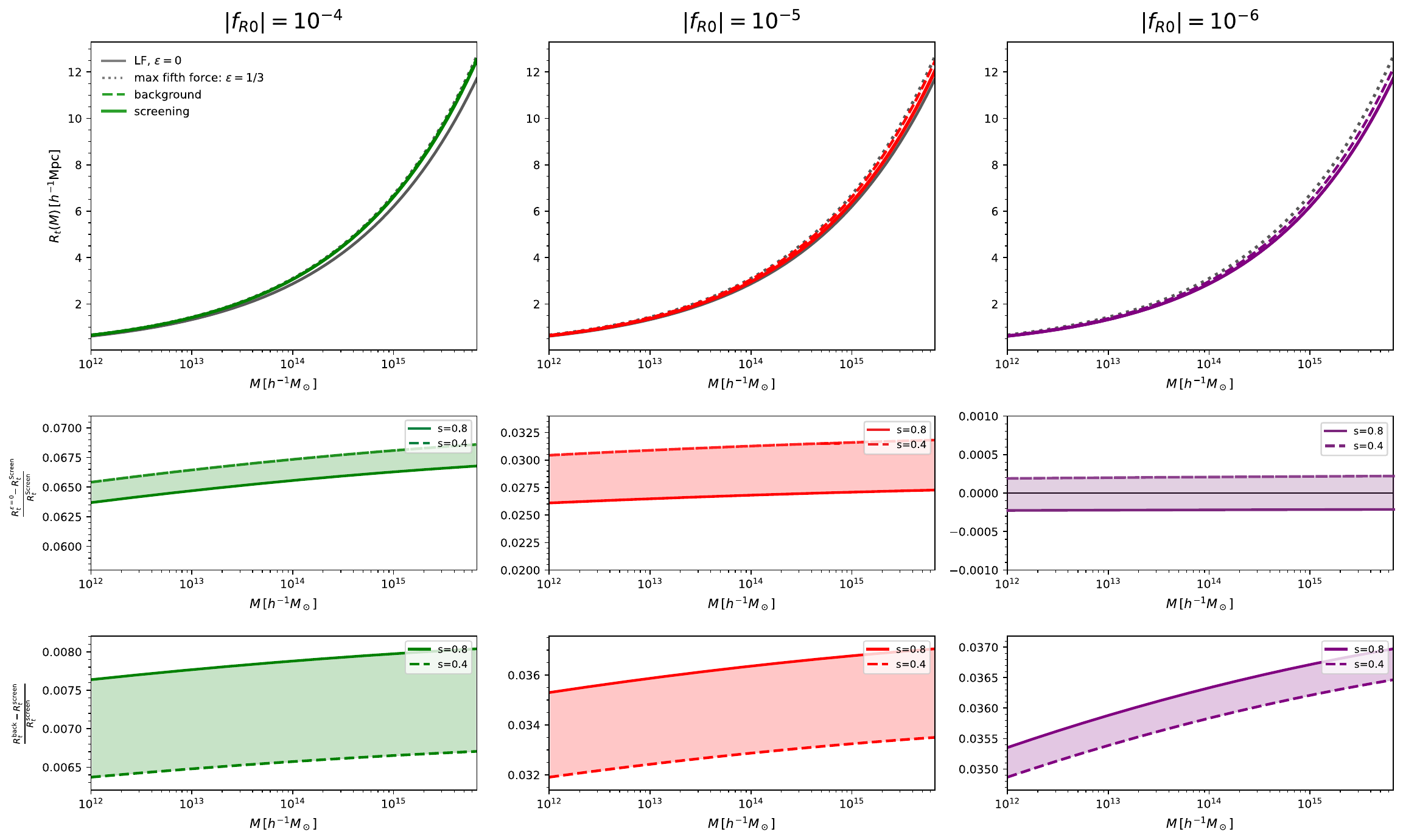}
\caption{Mass dependence of $R_t$. Symbols as in Fig. 4.}
\label{fig:Rt_M}
\end{figure*}

In the weak field case, $|f_{R0}| = 10^{-6}$, the scalar field is small and easily trapped by even low ambient matter densities. It screens almost immediately, exhibiting a smooth, continuous decay throughout cosmic history.
For a stronger field, $|f_{R0}| = 10^{-4}$,  the field has a high amplitude and requires the extreme densities found at much higher redshifts ($z > 1.5$) to be fully screened and forced into decay.

Fig. 8, bottom plot, represents more clearly the evolution of the background.   

Concerning the use of the turnaround radius to distinguish between GR and MG, we are 
in a similar situation to that of the previous figure.

The largest difference between the SF (GR) and the screened $R_t$ is $\simeq 6.5 \%$, in the case F4, but as we saw this case must be discarded from the results of the constraints. For F6, the "signal" is zero. We are left with the F5 case, in which the difference between GR and MG is $\simeq 3 \%$, not very high. We have to conclude as in the previous case: turnaround  is not very useful in distinguishing $f(R)$ theories from GR. 

In Fig. 9, we plot $R_t(M)$. Across the first row, the turnaround radius grows monotonically as the mass of the halo increases. Larger halos structurally support a much larger spatial radius.
The two bottom rows show similar changes as those shown in the previous two plots. 

For the strongest field (left), deviations step up smoothly over the mass range. For the weakest field (right), the fractional difference remains nearly flat and uniform, (zero value) across all halo masses because standard screening suppresses structural differences uniformly. 

In standard structure formation models, a virialized dark matter halo is defined as a bounded spherical overdensity. The total mass $M$ enclosed within a given structural boundary scales directly with the physical volume of the halo, this means that 
the physical radius, $R$, scales with halo mass, $M^{1/3}$. While the geometric baseline follows the $M^{1/3}$ law across all panels, the modified gravity field parameters ($|f_{R0}|$) introduce subtle shifts in the slopes between columns.
In the second row, for $|f_{R0}| = 10^{-6}$, massive halos ($M \sim 10^{15} \, h^{-1}M_{\odot}$) possess sufficiently deep gravitational potentials to trigger efficient local chameleon screening, forcing the curve to track standard GR, while
for $|f_{R0}| = 10^{-4}$ the scalar field environment remains largely unscreened even for massive clusters, altering the internal halo profile and slightly modifying the baseline structural slope.
Lower-mass halos ($M \sim 10^{12} \, h^{-1}M_{\odot}$) are highly abundant and typically have sufficient cosmic time to relax into symmetric, stable configurations. Conversely, cluster-scale halos ($M \sim 10^{15} \, h^{-1}M_{\odot}$) occupy the active nodes of the cosmic web and are dynamically young. They continuously experience violent major mergers, tidal disruptions, and asymmetric filamentary accretion. 
This ongoing dynamical assembly introduces massive object-to-object variations in their internal density profiles, subhalo distributions, and overall shape triaxiality. Because non-linear modified gravity metrics (such as the thin-shell thickness  corrections) depend heavily on the local density gradients, this intense structural diversity yields a wide spread in the calculated parameters, expanding the shaded scatter bands.

As in the previous plot, the percentage difference between $R_t$ in GR and the screened one, is  $6.5 \%$ for F4 and becomes 0 for F6. As before to distinguish between GR and $f(R)$ we have to consider F5, showing a small difference ($3 \%$). 

Summarizing, the idea that turnaround could successfully constrain cosmological parameters and explore the very nature of gravity introduced by \citep{Pavlidou2014,Faraoni2016}, and studied by several other authors (e.g., \citep{Bhattacharya2017,Lopes2018,Lopes2019,DelPopolo2020a,DelPopolo2020b}) is not viable. 

As already mentioned, the turnaround radius is not simply the spherical region obtained solving GR's equations, but it has a complex structure due to several physical effects. Moreover, a measurement of turnaround is complicated by the non-spherical nature of galaxy cluster environments, which instead consist of an intricate network of filaments, sheets, and voids. 
To this we must add that, as previously shown, as far as $f(R)$ theory is concerned one could use just F5, with discrepancies of a few percent with GR. 
In \citep{Hansen2020}, it was shown by means of numerical simulations that environmental asymmetries (such as filaments and cosmic voids) introduce a significant uncertainty margin in measurement precision, of the order of $20 \%$. This means that turnaround is not a good way at all to study the nature of gravity, at least in $f(R)$ models. This was already shown in \citep{DelPopolo2020a,DelPopolo2020b}. As shown by the same authors, the mass-temperature relation of clusters is not 
useful to disentangle between GR and $f(R)$ theories \citep{DelPopolo2019}, while the luminosity-temperature relation is a good way to  disentangle GR and $f(R)$ gravities \citep{DelPopolo2026}.  

\section{Conclusions}\label{sect:conclusions}

Interaction with external protostructures through shear and vorticity, and screening, modifies typical parameters of the SSCM. 
As already found in \citep{DelPopolo2013a,DelPopolo2013b,Pace2014,Mehrabi2017}, differently from the SSCM the collapse threshold $\delta_c$ depends on mass, redshift, and the parameter $f_{\rm R0}$ of the Hu-Sawicki $f_{\rm R}$ theory. 
Concerning $\delta_c$, the difference between the results for $\epsilon=0$ (GR), and $\epsilon=1/3$ (LF) obtained in the ESCM 
is of some percent, while the value of $\delta_c$ in GR in the ESCM, and the LF limit in the SSCM is much higher $\simeq 20 \%$. The behavior of $\delta_c(M)$, taking into account shear, vorticity, has the same dependence as the threshold for elliptical collapse obtained by \citep{DelPopolo1998,SMT2001,DelPopolo2017}. Its dependence on mass is twofold: dependence on mass through $f(R)$, and through shear and vorticity. 
By means of the excursion set approach, we obtained the mass function for F5 and F6 at $z=0$ (it can be easily found also at $z>0$). Even if there are no big differences with GR, it is possible to see a larger number of haloes in $f(R)$. This trend increases with increasing redshift, as shown by \citep{HernandezAguayo2018,Gupta2022,Gupta2022a}. 
In the case of $\delta_t$, the values are around $27$ and $29 \%$, considering ESCM or not, while for the turn-around, $R_t$, is around $6.5 \%$. Considering different values of $f_{\rm R0}$ in the case of $\delta_c$, the difference between GR value and the screened one is $2.5 \%$ for F4 and goes down to 0 for F6. $\delta_t$ shows larger differences with respect to $\delta_c$, a bit larger than $20 \%$ for F4. Recently, \citep{Artis2024} placed constraints on the values of  $f_{\rm R0}$ combining the overlapping Dark Energy Survey Year-3, KiloDegree Survey, and Hyper Suprime-Cam data for weak lensing mass calibration, and the first Spectrum Roentgen Gamma (SRG)/eROSITA All-Sky Survey (eRASS1) cluster catalog in the western Galactic hemisphere. Assuming massless neutrinos it was shown that $|f_{\rm R0}|<10^{-4.31}$, at $95 \%$  confidence level, and $|f_{R0}| \lesssim 10^{-4.5}$ (at large scales), in other studies. This excludes F4. In order to disentangle between GR and $f(R)$, only F5 could be used, since F4 is excluded, $F6$ is practically equal to GR, and the LF is excluded because it is larger than F4. 

Unfortunately, as shown by some studies, \citep{Hansen2020}, the error on $R_t$, is of the order of $20\%$, excluding the possibility to use the turnaround radius, and related quantities, to disentangle between GR and $f(R)$. 

In order to disentangle between GR and $f(R)$, we are left with the mass function which is a good probe especially at increasingly high redshifts. 

\begin{acknowledgments}

The author would like to thank Niayesh Afshordi for useful discussions, and is grateful for the hospitality of Perimeter Institute where this work was carried out. Research at Perimeter Institute is supported in part by the Government of Canada through the Department of Innovation, Science and Economic Development Canada and by the Province of Ontario through the Ministry of Colleges, Universities, Research Excellence and Security.
\end{acknowledgments}

\newpage

\appendix

\section{f(R) theories and the Hu-Sawicki model }\label{sect:MG}

In this section, we summarize the MG theory and in particular the Hu-Sawicki model. 
The Hu-Sawicki model is a particular model among the $f(R)$ theories. 

An MG model such as $f(R)$ can be obtained by modifying the Einstein--Hilbert action, replacing the Ricci scalar $R$ with $R+f(R)$, as follows: 

\begin{equation}
S =  \int{d^4 x \sqrt{-g} \left[ \frac{R+f(R)}{16\pi G} + \mathcal{L}_{m} \right]} \, ,\label{eqn:action}
\end{equation}
where $\mathcal{L}_{m}$ is the matter Lagrangian density. The $\Lambda$CDM model is recovered in the case $f(R)= -2 \Lambda$. 

Varying the action, the modified Einstein equations are given by:
\begin{equation}
G_{\alpha \beta} + f_R R_{\alpha \beta} - \left(\frac{f}{2} - \Box f_R \right)g_{\alpha \beta}- \nabla_{\alpha} \nabla_{\beta} f_R = 8\pi G \, T_{\alpha \beta} \, , \label{egm}
\end{equation}
where $G_{\alpha \beta}$ is the Einstein tensor, $f_{R} \equiv \frac{d f(R)}{dR}$, and $T_{\alpha \beta}$ is the energy-momentum tensor. 
Through the trace of Einstein's modified equations, one can obtain a Klein-Gordon equation for $f_R$, which can be considered
as a scalar field. Field and metric are coupled through: 
\begin{equation}
3\Box f_R-R+f_R R-2f=-8\pi G T, 
\end{equation}
where $T$ is the trace of the energy-momentum tensor. The scalaron effective potential may be defined through
\begin{equation}
\frac{\partial V_{\rm eff}}{\partial f_R}
=
\frac{1}{3}\left(R-f_R R+2f-8\pi G T\right),
\end{equation}
and its mass is
\begin{equation}
m_{f_R}^2
=
\frac{\partial^2 V_{\rm eff}}{\partial f_R^2}
=
\frac{1}{3}
\left[
\frac{1+f_R}{f_{RR}}-R
\right].
\end{equation}

In viable high-curvature models, $|f_R|\ll 1$ and $\qquad |R f_{RR}|\ll 1$ 
so that
\begin{equation} \label{eq:mf}
m_{f_R}^2\simeq \frac{1}{3f_{RR}}.
\end{equation}

As seen, in $f(R)$ gravity, the Einstein--Hilbert action is generalized by replacing the Ricci scalar
with a nonlinear function of the curvature. This modification introduces an additional scalar
degree of freedom, usually referred to as the scalaron, whose dynamics are governed by a
Klein--Gordon equation derived from the trace of the modified Einstein equations. The mass
of this scalar field determines the range of the fifth force and therefore controls the
deviations from GR.
The inverse of the scalaron mass defines the comoving Compton wavelength,
\begin{equation}
\lambda_{\rm C}=\frac{1}{m_{f_R}},
\end{equation}
which determines the characteristic range over which the additional scalar degree of freedom
mediates gravitational interactions. Since the scalar field couples to matter through the trace
of the energy--momentum tensor, it generates an extra attractive interaction, commonly
referred to as a \emph{fifth force}, in addition to the standard Newtonian gravitational force.

The strength and range of this force depend on the scalaron mass. A light scalar field
corresponds to a large Compton wavelength, allowing the fifth force to operate over
cosmological distances. Conversely, when the scalaron becomes massive, the interaction range
is strongly suppressed and the theory approaches the predictions of GR.

In viable $f(R)$ models, the scalaron mass increases in high-density environments through the
\emph{chameleon mechanism}. This environmental dependence efficiently screens the fifth
force inside dense astrophysical systems, such as stars, galaxies, and the Solar System,
thereby satisfying the stringent local tests of gravity while still allowing detectable
deviations from GR on cosmological scales.

In the quasi-static regime (subhorizon limit), perturbing around the cosmological background gives
\begin{equation}
\nabla^2\delta f_R
=
\frac{a^2}{3}
\left(
\delta R-8\pi G\delta\rho_m
\right),
\end{equation}
where $\delta f_R = f_R(R)-f_R(\bar{R})$, $\delta R = R-\bar{R}$, and $\delta\rho_m=\rho_m-\bar{\rho}_m$,
where an overbar denotes the homogeneous cosmological background.

The modified Poisson equation is
\begin{equation}
\nabla^2\Phi
=
\frac{16\pi G}{3}a^2\delta\rho_m
-\frac{a^2}{6}\delta R .
\end{equation}

If the scalar-field perturbation is linearized,
\begin{equation}
\delta R\simeq
\frac{dR}{df_R}\delta f_R ,
\end{equation}
one obtains in Fourier space
\begin{equation}
-k^2\Phi(k,a)
=
4\pi G a^2
\left[1+\epsilon(k,a)\right]
\delta\rho_m(k,a),
\end{equation}
with
\begin{equation} \label{eq:eps_appendix}
\epsilon(k,a)
=
\frac{1}{3}
\frac{k^2}{a^2m_{f_R}^2+k^2}.
\end{equation}

Thus the effective Newton constant is
\begin{equation}
G_{\rm eff}(k,a)
=
G\left[1+\epsilon(k,a)\right].
\end{equation}

The GR limit is recovered when
\begin{equation}
a^2m_{f_R}^2\gg k^2,
\qquad
\epsilon(k,a)\rightarrow 0,
\end{equation}
the so-called small field (SF) limit,
whereas the unscreened large-field limit gives
\begin{equation}
a^2m_{f_R}^2\ll k^2,
\qquad
\epsilon(k,a)\rightarrow \frac{1}{3},
\end{equation}
which is the large field (LF) limit. 

\subsection{Hu--Sawicki model}

The Hu--Sawicki model can be written as
\begin{equation} \label{eq:model}
f(R)
=
-m^2
\frac{c_1(R/m^2)^n}
{c_2(R/m^2)^n+1},
\end{equation}
where $m$, $c_1$, $c_2$, and $n$, (with $n>0$) are free parameters, with  $m^2 = H_0^2\Omega_{\rm m,0}$, while $c_1$ and $c_2$ can be determined through
\begin{equation}
 \frac{c_1}{c_2} \approx 6\frac{\Omega_{\Lambda,0}}{\Omega_{\rm m,0}}\,.
\end{equation}
As seen, in the large curvature regime ($R/m^2\gg 1$), $f(R)\approx -2\Lambda$. 

In the high-curvature regime and for \(n=1\), expanding Eq. \ref{eq:model}
\begin{equation}
f(R)\simeq
-16\pi G\rho_\Lambda
-f_{R0}\frac{\bar{R}_0^2}{R}.
\end{equation}

where the leading constant term has been matched to a cosmological constant,
$\bar{R}_0=\bar{R}(z=0)$ and $f_{R0}=f_R(\bar{R}_0)$.
The standard $\Lambda$CDM model is recovered in the limit $|f_{R0}| \to 0$.

{We stress that, in this ``designer'' Hu-Sawicki construction, $f(R)$ is not intended to \emph{replace} $\Lambda$ as an alternative explanation of the dark-energy phenomenon: the ratio $c_1/c_2$ is tuned precisely so that the high-curvature/background limit of $f(R)$ reproduces $-2\Lambda$, i.e., the background expansion history is built by construction to mimic $\Lambda$CDM. $\Omega_{\Lambda,0}$ is therefore retained throughout, both in the background expansion rate $E(a)$ (Eq. \ref{eq:e}) and, as shown below, in the scalaron-mass expression, while all of the new, testable physics of the model is carried by the single deviation parameter $f_{R0}$, which controls the scalar (chameleon) sector governing the fifth force and its screening. Consequently, $\Omega_{\Lambda,0}$ enters numerically in every calculation presented in this paper: through $E(a)$ in the master evolution equation for $\delta$ (Eq. \ref{eq:HuSaw}), and through the scalaron mass used to compute the screening term $\epsilon(k,a)$ (Eq. \ref{eq:eps}) at every radius and epoch.}

We may also rewrite the mass of the scalar field in Eq. \ref{eq:mf}
as:
\begin{equation}
m_{f_R}
\simeq
\frac{H_0}{c}
\left(
\frac{\Omega_{m0}+4\Omega_\Lambda}{2|f_{R0}|}
\right)^{1/2}
\left[
\frac{\Omega_{m0}(1+z)^3+4\Omega_\Lambda}
{\Omega_{m0}+4\Omega_\Lambda}
\right]^{3/2}.
\end{equation}

Through the equations presented in the previous section, we may investigate the evolution of matter density perturbations within the Hu--Sawicki $f(R)$ framework.
Unless otherwise stated, we adopt the fiducial cosmological parameters
$\Omega_m=0.31$, $\Omega_\Lambda=0.69$, $\sigma_8=0.81$, and $n_s=0.96$ 

\section{The ESCM}

We need to calculate the evolution equations of $\delta$ in the non-linear regime, already obtained by \cite{Bernardeau1994,Padmanabhan1996,Ohta2003,Ohta2004,Abramo2007,Pace2010} in the spherical and ellipsoidal collapse models. To start with, for the fluid we assume the equation of state $P=w\rho c^2$, and the Neo-Newtonian expressions for the relativistic Poisson equation, the Euler, and continuity equations \citep{Lima1997}
\begin{eqnarray}
  \frac{\partial\rho}{\partial t}+\nabla_{\vec{r}}\cdot(\rho\vec{v})+
  \frac{P}{c^2}\nabla_{\vec{r}}\cdot\vec{v} & = & 0 \label{eqn:cnpert}\;,\\
  \frac{\partial\vec{v}}{\partial
    t}+(\vec{v}\cdot\nabla_{\vec{r}})\vec{v}+
  \nabla_{\vec{r}}\Phi +{\bf \frac{c^2}{c^2 \rho+P} \nabla P} & = & 0\;, \label{eqn:enpert}\\
  \nabla^2\Phi-4\pi G\left(\rho+\frac{3P}{c^2}\right) & = & 0\;,\label{eqn:pnpert}
\end{eqnarray}
Here, $\vec{r}$ the physical coordinate, $\Phi$ the Newtonian gravitational potential, and  $\vec v$ the velocity in three-space.

Let's use comoving coordinates $\vec{x}=\vec{r}/a$ to obtain the perturbations equations

\begin{eqnarray}
 \dot{\delta}+(1+w)(1+\delta)\nabla_{\vec{x}}\cdot\vec{u} & = & 0\;, \label{eq:pertCont}\\
 \frac{\partial \vec{u}}{\partial t}+2H\vec{u}+(\vec{u}\cdot\nabla_{\vec{x}})\vec{u}+\frac{1}{a^2}\nabla_{\vec{x}}\phi &
 = & 0\;,\label{eq:pertEuler} \\
 \nabla_{\vec{x}}^2\phi-4\pi G(1+3w)a^2\bar{\rho}\delta & = & 0\;, \label{eq:pertPois}
\end{eqnarray}
being $\vec{u}(\vec{x},t)$ is the comoving peculiar velocity, and $H(a)$ the Hubble function 
\footnote{The pressure terms in Eq. (\ref{eq:pertEuler})  are zero, since in the top-hat, with $\rho=\overline{\rho} +\delta \rho$, and $P=\overline{P} +\delta P$ the gradient is zero.}

The non-linear evolution equation is obtained combining the previous equations.
In the case $w=0$ (dust), we get:
\begin{equation}\label{eqn:nleq11}
\begin{split}
\ddot{\delta}+2H\dot{\delta}-\frac{4}{3}\frac{\dot{\delta}^2}{1+\delta}-
4\pi G\bar{\rho}\delta(1+\delta)-\\
(1+\delta)(\sigma^2-\omega^2) & = 0\;
\end{split}
\end{equation}
Eq. (\ref{eqn:nleq11}) is Eq.~41 of \cite{Ohta2003}, and a generalization of Eq.~7 of \cite{Abramo2007} to the case of a non-spherical configuration of a rotating fluid.

In Eq. (\ref{eqn:nleq11}), $\sigma^2=\sigma_{ij}\sigma^{ij}$, is the shear, related to a symmetric traceless tensor, named shear tensor and $\omega^2=\omega_{ij}\omega^{ij}$ is the vorticity, related to an antisymmetric tensor. They are given by:
\begin{eqnarray}
 \sigma_{ij} & = & \frac{1}{2}\left(\frac{\partial u^j}{\partial x^i}+\frac{\partial u^i}{\partial x^j}\right)
 -\frac{1}{3}\theta\delta_{ij}\;, \\
 \omega_{ij} & = & \frac{1}{2}\left(\frac{\partial u^j}{\partial x^i}-\frac{\partial u^i}{\partial x^j}\right)\;.
\label{eq:tens}
\end{eqnarray}
where $\theta=\nabla_{\vec{x}}\cdot\vec{u}$ is the expansion.

In terms of the scale factor $a$, Eq. (\ref{eqn:nleq11}) takes the form:

\begin{equation}\label{eqn:wnldeq}
 \begin{split}
  \delta^{\prime\prime}+\left(\frac{3}{a}+\frac{E^\prime}{E} \right)
\delta^\prime-\frac{4}{3}\frac{\delta^{\prime 2}}{1+\delta}-
  \frac{3}{2}\frac{\Omega_{\mathrm{m},0}}{a^5 E^2(a)}\delta(1+\delta)-&\\
  \frac{1}{a^2H^2(a)}(1+\delta)(\sigma^2-\omega^2)&=0\;,
 \end{split}
\end{equation}
where $\Omega_{m,0}$ is DM density parameter today ($a=1$).

The term $E(a)$ is given by
\begin{equation}\label{eq:e}
E(a)=\sqrt{\frac{\Omega_{\mathrm{m},0}}{a^3}+\frac{\Omega_{\mathrm{K},0}}{a^2}+ \Omega_{\mathrm{\Lambda},0}g(a)}\;,
\end{equation}
and  
\begin{equation}
g(a)=\exp{\left(-3\int_1^a\frac{1+w(a')}{a'}~da^{\prime}\right)}\;.
\end{equation}

We need an expression for $\sigma^{2} - \omega^{2}$. 
As shown in \citep{DelPopolo2013a,DelPopolo2013b,Pace2014,Mehrabi2017}, 
using a spherical collapse which considers angular momentum, we consider the ratio between the rotational term and the gravitational one, obtaining $\alpha=\frac{L^2}{M^3 R G} \simeq \frac{R V^2}{M  G}$.

{We stress that $\alpha$ encodes a genuine physical mechanism, namely angular-momentum acquisition via tidal torquing from the surrounding large-scale structure (``protostructures'') during collapse, in the standard tidal-torque-theory picture, rather than a purely effective device introduced only to allow shear and vorticity terms in the model. This is reflected in the fact that $\alpha$ is not tuned to reproduce this paper's own results, but is calibrated directly against independent, observed rotation-velocity data for the Milky Way, the Coma cluster (optical and XRISM), and Abell 2107, as summarized below. We note, however, that other physical channels not currently included in this parameterization -- in particular the highly nonlinear dynamics of dark-matter self-clustering during virialization, and the interaction with internally accreting baryonic gas -- may also contribute to shear and vorticity, especially at late times and on small scales, and represent a natural, complementary extension of the present framework that we plan to pursue in future work.}

The above quoted ratio, $\alpha=\frac{L^2}{M^3 R G}$, in the case of the Milky Way using \citep{Sofue2020} data  is $ \simeq 0.08$. 

In \citep{DelPopolo2013}, we found a similar value, 0.05, and 0.04 is given also by \citep{Marciu2016}. Then at $10^{12} M_{\odot}$, $\alpha$ should be in the range 0.05-0.08.

The value of $\alpha$ is larger for smaller size perturbations (dwarf-galaxy-size size perturbations) and smaller for larger 
size perturbations (clusters). 

Concerning clusters, e.g. Coma, several estimates of the coherent rotation velocity are available:
the historical optical measurement $v_{rot}\simeq 137~\mathrm{km/s}$ \citep{Colless1996} (marginal at $\sim\!1.3\sigma$), and the recent XRISM detection of a North--South velocity gradient of the intracluster
gas of $\sim\!530~\mathrm{km/s}$ across the cluster, corresponding to
$v_{rot}\simeq265~\mathrm{km/s}$ per side ($530~\mathrm{km/s}$ being an
upper limit). The corresponding values of
$\alpha=v_{rot}^{2}R/(GM_{tot})$ are:
\begin{align}
v_{rot}=137~\mathrm{km/s} &\;\Rightarrow\; \alpha \simeq 6.7\times10^{-3}\,,\\
v_{rot}=265~\mathrm{km/s} &\;\Rightarrow\; \alpha \simeq 2.5\times10^{-2}\,.
\end{align}

Abell 2107 is the best-studied rotating cluster, with a robust
rotation detection \citep{Song2018}. With \citep{Song2018} data $v_{rot}=160\text{--}208~\mathrm{km/s}$ (at $3.5\sigma$), and then yielding $\alpha=0.021$--$0.035$.

The most reliable cluster-scale estimate is therefore in the range $0.0067-0.035$ with an average of $\alpha \sim10^{-2}$ (Coma marginal detection and A2107 outer-region
robust detection), i.e.\ roughly one order of magnitude below the galactic value.
Also at cluster scales, \citep{Mehrabi2017}, performing a likelihood analysis with current available observational data,
found a  value of $\simeq 0.005$ at $10^{15} M_{\odot}$. 

Using the same $ansatz$ we may calculate the same ratio between the gravitational and the extra term appearing 
in Eq.~\ref{eqn:wnldeq} thereby obtaining  
\begin{equation} \label{eq:sigm}
(\sigma^2-\omega^2)H_0^{-2}= -\frac{3}{2} \frac{\alpha\Omega_{\mathrm{m},0}}{a^3}\delta.  
\end{equation}

Then substituting Eq. \ref{eq:sigm} in Eq. \ref{eqn:wnldeq} the nonlinear equation to solve is 

\begin{equation}\label{eqn:wnldeqq}
  \delta^{\prime\prime}+\left(\frac{3}{a}+\frac{E^\prime}{E} \right)
\delta^\prime-\frac{4}{3}\frac{\delta^{\prime 2}}{1+\delta}-
  \frac{3}{2}\frac{\Omega_{\mathrm{m},0} (1-\alpha)}{a^5 E^2(a)}\delta(1+\delta)=0.
\end{equation}

In order to obtain the term $\alpha(M)$, in the range previously considered 
($10^{12} h^{-1} M_\odot$ - $6.7 \times 10^{15} h^{-1}M_\odot$), we 
calculated the threshold of collapse $\delta_c$, with our model, and matched it with $\delta_c$ obtained by Del Popolo \citep{DelPopolo1999,DelPopolo2017}, or Sheth, Mo, and Tormen \citep{SMT2001}. The result obtained, is plotted in Fig. 5b. 
$\alpha(M)$ is fitted by the quartic function of $\sigma(M)$ (written on the panel of Fig. 5b), where $\sigma(M)$ 
the root-mean-square density fluctuation on a mass scale $M$ given in Eq. B6 of \citep{DelPopolo2017}.

The value of $\alpha(M)$ is 0.05925 at $10^{12}\,h^{-1}M_\odot$, $\simeq0.01431$ at $10^{15}\,h^{-1}M_\odot$, and $\simeq0.00779$ at $6.7\times10^{15}\,h^{-1}M_\odot$, in agreement with the values estimated above.

The threshold of collapse, and the turnaround, can be obtained solving Eq. (\ref{eqn:wnldeqq}) following the method described in \citep{Pace2010}. 

\section{Inclusion of the $f(R)$ modification}

Eq. \ref{eqn:wnldeq}, in terms of the potential $\Phi$ may be written as

\begin{equation}\label{eq:imp}
\begin{split}
\delta'' &+ \left( \frac{3}{a} + \frac{E'}{E} \right) \delta' - \frac{4}{3} \frac{\delta'^2}{1 + \delta} \\
&- \frac{1}{a^2 H^2(a)} (1 + \delta)(\sigma^2 - \omega^2) = \frac{1 + \delta}{H^2 a^4} \nabla^2 \Phi 
\end{split}
\end{equation}

In the Hu--Sawicki model, after reducing to the quasi-static regime and linearising in $\delta f_R$,
the modified Poisson equation in Fourier space reads [\citep{Lopes2018}, Eq. (2.9)] 

\begin{equation} \label{eq:poiss}
-k^2 \Phi(k,a)
=
[1+\epsilon(k,a)]\,4\pi G a^2\bar\rho\,\delta(k,a),
\end{equation}
with $\epsilon(k,a)$ given by Eq. \ref{eq:eps_appendix}. In configuration space, inverting Eq. \ref{eq:poiss} with the convention
\[
\Phi(\mathbf x,a)
=
\int\frac{d^3k}{(2\pi)^3}
e^{i\mathbf k\cdot\mathbf x}
\Phi(k,a)
\]
gives
\begin{equation}
\Phi(\mathbf x,a)
=
-4\pi G a^2\bar\rho
\int\frac{d^3k}{(2\pi)^3}
e^{i\mathbf k\cdot\mathbf x}
k^{-2}
[1+\epsilon(k,a)]\delta(k,a),
\end{equation}
in agreement with Eq. (3.2) of \citep{Lopes2018}. Applying $\nabla^2$, which acts as $-k^2$
in Fourier space, cancels the $k^{-2}$:
\begin{equation}
\nabla^2\Phi(\mathbf x,a)
=
4\pi G a^2\bar\rho
\int\frac{d^3k}{(2\pi)^3}
e^{i\mathbf k\cdot\mathbf x}
[1+\epsilon(k,a)]\delta(k,a).
\end{equation}
Under spherical symmetry the angular integral gives the standard reduction
\begin{equation}
\int\frac{d^3k}{(2\pi)^3}
e^{i\mathbf k\cdot\mathbf x}
f(k)
=
\frac{1}{2\pi^2}
\int_0^\infty dk\,k^2 f(k)
\frac{\sin(kr)}{kr},
\end{equation}
so that

\begin{equation}\label{eq:nablo}
\begin{split}
\frac{1}{a^2} \nabla^2 \Phi(r,a) 
&= \frac{4\pi G\bar{\rho}}{2\pi^2} \int_0^\infty dk \, k^2 [1 + \epsilon(k,a)] \delta(k,a) \frac{\sin(kr)}{kr} \\
&= \frac{2G\bar{\rho}}{\pi} \int_0^\infty dk \, k^2 [1 + \epsilon(k,a)] \delta(k,a) \frac{\sin(kr)}{kr}.
\end{split}
\end{equation}

Using
\[
4\pi G\bar\rho=\frac{3}{2}H_0^2\Omega_{m,0}a^{-3},
\qquad
\frac{2G\bar\rho}{\pi}
=
\frac{3H_0^2\Omega_{m,0}}{4\pi^2a^3},
\]
Eq. \ref{eq:nablo} reads
\begin{equation}
\frac{1}{a^2}\nabla^2\Phi(r,a)
=
\frac{3H_0^2\Omega_{m,0}}{4\pi^2a^3}
\int_0^\infty dk\,k^2
[1+\epsilon(k,a)]\delta(k,a)
\frac{\sin(kr)}{kr}.
\end{equation}

Substituting the previous equations in Eq. 
\ref{eq:imp}, 
we get

\begin{equation}
\begin{split}
\delta''+
\left(\frac{3}{a}+\frac{E'}{E}\right)\delta'
-\frac{4}{3}\frac{\delta'^2}{1+\delta}
-\frac{1+\delta}{a^2H_0^2E^2}
(\sigma^2-\omega^2)
\\
=
\frac{3(1+\delta)\Omega_{m,0}}
{4\pi^2a^5E^2}
\int_0^\infty
dk\,k^2
\left[1+\epsilon(k,a)\right]
\delta(k,a)
\frac{\sin(kr)}{kr}.
\end{split}
\label{eq:collapse_shear}
\end{equation}

which is the generalization of Eq. (3.3) of \citep{Lopes2018} to include shear and
vorticity. Using Eq. \ref{eq:sigm}
\begin{equation} \label{eq:evol}
\begin{aligned}
&\delta''+
\left(\frac{3}{a}+\frac{E'}{E}\right)\delta'
-\frac{4}{3}\frac{\delta'^2}{1+\delta}
= \\[2mm]
&\frac{3(1+\delta)\Omega_{m,0}}
{2a^{5}E^{2}}
\left[
\frac{1}{2\pi^{2}}
\int_{0}^{\infty}
dk\,k^{2}
\left[1+\epsilon(k,a)\right]
\delta(k,a)
\frac{\sin(kr)}{kr}
-\alpha \delta
\right].
\end{aligned}
\end{equation}
which is the same equation written in terms of $\alpha \equiv \alpha(M)$.

\section{Chameleon mechanism}

The scalar field has an environment-dependent behaviour which was first investigated by \citep{Khoury2004,Mota2007} and is often referred to as the ‘chameleon effect’. In cosmology and modified gravity theories (such as the Hu-Sawicki $f(R)$ model), 
the chameleon mechanism consists in a scalar field whose mass, and hence the range of the fifth force it mediates, depends on the local matter density.

In the range of mass we study one has to be cautious and modify the ESCM to take the mechanism into account. Then the chameleon mechanism is a screening effect where the scalar field responsible for the "fifth force" (the enhanced gravity,  e.g. $\epsilon = 1/3$) acquires a mass that depends on the local ambient density. For small masses ($< 10^{11} M_{\odot}$) gravitational potential wells are shallow, and the chameleon mechanism does not activate (unscreened regime). Modified gravity is fully active, making collapse easier. For masses $>10^{16} M_{\odot}$ potential wells are extremely deep (massive galaxy clusters). The chameleon is fully activated (screened regime). The fifth force is suppressed, and the system reverts entirely to standard GR ($\epsilon = 0$). 
So, it is paramount to take it into account. 

In the evolution equation of \citep{Lopes2018}, only the term 
\begin{equation}
\epsilon(k,a)=
\frac13
\frac{k^2}{a^2m_{f_R}^2(a)+k^2}.
\end{equation}
appears.

The fundamental limitation of their approach is that $\epsilon(k,a)$ depends only on the Fourier scale and on the background scalaron mass, never on the local density of the collapsing region. Consequently, the original model neglects the chameleon screening mechanism.

It is necessary, somehow, to introduce it. There are several ways.
In Hu–Sawicki $f(R)$ gravity, the Einstein equations become:
\begin{equation}
G_{\mu\nu} = 8\pi G T_{\mu\nu} + T_{\mu\nu}^{\text{(sf)}}
\end{equation}
Under the quasi-static approximation, the system reduces to three coupled equations:
\begin{equation}
\nabla^2 \Phi = 4\pi G a^2 \bar\rho_m \delta - \frac{1}{2} \nabla^2 \delta f_R
\end{equation}

\begin{equation} \label{eq:fr}
\nabla^2 \delta f_R = \frac{a^2}{3} \left[ \delta R(f_R) - 8\pi G \delta \rho_m \right]
\end{equation}

and the Euler (fluid) equation governing the motion of collisionless matter 

\begin{equation}
\ddot{\mathbf{x}}
+
2H\dot{\mathbf{x}}
=
-\frac{1}{a^2}\nabla\Phi ,
\label{eq:Euler}
\end{equation}

where $\mathbf{x}$ denotes the comoving coordinate and $H=\dot a/a$ is the Hubble parameter.

For spherical symmetry, Eq.~(\ref{eq:Euler}) reduces to

\begin{equation}
\ddot r
=
-\frac{\partial\Phi}{\partial r}
+
\frac{\ddot a}{a}r,
\label{eq:radial}
\end{equation}
where $r=a\,x$ is the physical radius of the spherical shell.
Solving the equations, one can calculate the collapse of a spherically symmetric, pressureless matter distribution in an
asymptotic FRW spacetime. While in GR, due to Birkhoff's theorem an initial homogeneous overdensity ("top-hat") does not change its shape; this is no longer true 
in $f(R)$ theories, since Birkhoff's theorem is no longer valid due to the existence of a further scalar degree of freedom allowing monopole radiation. The spherical region is divided into many concentric shells; Eq. (\ref{eq:fr}) is solved to obtain the fifth force, which, together with a collapse condition, allows one to determine the needed quantities.

This approach was followed by \citep{Kopp2013} to get the collapse threshold $\delta_c(M,z,f_{\rm R0})$, and the mass function. The approach is computationally expensive; moreover they, like all others papers trying to get $\delta_c$, skipped 
shear and vorticity. The effect of this is that the predicted $\delta_c$ (see Fig. 6 in \citep{Kopp2013}) are monotonically 
increasing with mass up to the GR limit.

In other words, \citep{Kopp2013} and related papers used a spherical collapse model which, by its nature, does not take into account the interaction of a perturbation with the external ones.

In other words, in \citep{Kopp2013}, the mass dependence of $\delta_c$ comes from the MG. In reality, mass dependence must appear in the collapse even in GR as shown by \citep{DelPopolo1998a,SMT2001}. 
This implies that the resulting mass function is not fully correct.

Several papers used simpler approaches with respect to that of \citep{Kopp2013}, e.g. \citep{Borisov2012,Li2012,Lam2012,Cataneo2016}, some of which used the dynamical thin shell approximation. 

Other simpler approaches starts from the observation that in the standard linearized background approach, the scalaron mass $m_{f_R}$ depends strictly on the cosmological background scale factor $a$. To capture the authentic chameleon effect, where gravity reverts to GR inside dense regions, one must promote $m_{f_R}$ to depend dynamically on the local density $\rho_m = \bar{\rho}_m(1 + \delta)$:

$$
\begin{aligned}
m_{f_R}(\delta, a) \simeq & \frac{H_0}{c} \left[ \frac{\Omega_{m0} + 4\Omega_\Lambda}{2|f_{R0}|} \right]^{1/2} \\
& \times \left[ \frac{\Omega_{m0}(1+z)^3(1+\delta) + 4\Omega_\Lambda}{\Omega_{m0} + 4\Omega_\Lambda} \right]^{3/2}
\end{aligned}
$$

Consequently, the scale-dependent modification filter $\varepsilon$ becomes an explicit function of the local overdensity $\delta$:

\begin{equation}
\varepsilon(k, a, \delta) = \frac{1}{3} \frac{k^2}{a^2 m_{f_R}^2(\delta, a) + k^2}
\end{equation}

As the collapse progresses and $\delta$ grows, $m_{f_R}^2 \propto (1 + \delta)^3$ escalates rapidly. This drives $\varepsilon \to 0$, shutting off the fifth force ($G_{\text{eff}} \to G$) inside the overdensity.
Since $\varepsilon$ now varies point-by-point based on $\delta(r, a)$, it can no longer be processed as a uniform global multiplier in Fourier ($k$) space. One can resolve this issue using a spatial representative value of $\delta$ for the entire collapsing boundary 
to update $m_{f_R}(a, \bar{\delta}(a))$ globally at each time step. This approach has several drawbacks. 

When using a uniform, single-value representation of the overdensity $\delta$ to globally update the scalaron mass $(m_{f_{R}}$) at each time step, one introduces several major physical and mathematical limitations: A. Unphysical Uniformity: it forces the fifth force to "turn off" simultaneously across the entire boundary or halo, failing to capture the continuous shielding and spatial variation of the chameleon field. B. Mass underestimation: using central peak densities artificially elevates the global mass $m_{f_{R}}$, which suppresses structure growth more than is physically accurate for the outskirts of the halo. Conversely, volume averages underestimate the mass in the dense core. C. Violation of locality: the chameleon mechanism is inherently local. Enforcing a global filter dampens the true scale-dependence of screening, which requires high-mass (screened) centers and low-mass (unscreened) outskirts.

A rigorous treatment, already mentioned and adopted by \citep{Kopp2013}, is to explicitly solve the linearized Klein--Gordon equation pointwise:

    \begin{equation}
    \nabla^2 \delta f_R = \frac{a^2}{3}(\delta R - 8\pi G \delta \rho_m)
    \end{equation}
which requires integrating an elliptic relaxation or multigrid solver at every time step, or a N-body code (e.g., ECOSMOG, MG-GADGET). 

An intermediate approach, not very rigorous but better than the one discussed before, 
consists in promoting the gravitational modification factor from a purely scale-dependent quantity to one that also depends on the local density profile through a dynamical screening function,i.e.\ local screening via the thin-shell condition. 
The modified gravity correction therefore becomes

\begin{equation}
\epsilon(k,a)\longrightarrow\epsilon(k,a)\,S(r,a),
\end{equation}

where $0\le S(r,a)\le1$.
The screening factor satisfies the Khoury-Weltman condition:

\begin{equation} \label{eq:KW}
S(r,a)=
\min\left[
1,
3\frac{\Delta R_{\rm th}(r,a)}{R(r,a)}
\right],
\end{equation}

where

\begin{equation}
\frac{\Delta R_{\rm th}(r,a)}{R(r,a)}
=
\frac{\bar f_R(a)-f_{R,\min}(\rho_{\rm in}(r,a))}
{6\Phi_N(r,a)}.
\end{equation}

There are several factors entering the screening factor such as: 1. the Newtonian potential, $\Phi_N(r,a) = \frac{GM(<r)}{a\,r}$, of the mass enclosed within $r$; 2. the local equilibrium value of the scalar field for the instantaneous internal density $\rho_{\rm in} (r,a)$, $f_{R,\min}(\rho_{\rm in}) = -f_{R0} \frac{\bar R_0^2}{R(\rho_{\rm in})}$, 3. the cosmological background, $\bar f_R(a)$.
Unlike the original formulation, the screening factor depends on the evolving density profile through the enclosed mass, the Newtonian potential, and the local density. Consequently, the collapse equation becomes genuinely nonlinear and self-consistent.

A limitation of this approach is that replacing the nonlinear elliptic equation

\[
\nabla^2 \delta f_R
=
\frac{a^2}{3}
\left[
\delta R(f_R)-8\pi G\delta\rho
\right].
\]

with an algebraic thin-shell prescription neglects the finite propagation of the scalar field, the radial gradients, and the
non-local coupling between neighboring shells.

Nevertheless, the one-zone approximation captures the dominant physical mechanism responsible for chameleon screening while requiring only minimal modifications to the original \citep{Lopes2018} numerical solver.

The natural extension of the one-zone approximation is to retain the full
radial density profile while introducing the chameleon mechanism locally,
shell by shell, rather than through a single global screening factor. The
evolving configuration is represented by the profile $\delta(r,a)$, sampled
on the radial grid of comoving radii $r_1,\ldots,r_N$. For each shell $r_i$, the enclosed mass
is
\begin{equation}
M(<r_i) = 4\pi \int_0^{r_i} \rho(r)\,a^3 r^2\,dr,
\end{equation}
or, numerically,
\begin{equation}
M_i = 4\pi a^3 \sum_{j=1}^{i} \rho_j\, r_j^2\, \Delta r .
\end{equation}

This gives a \emph{local}, shell-dependent Newtonian potential
\begin{equation}
\Phi_N(r_i) = \frac{G\,M_i}{a\,r_i},
\end{equation}
in place of the single global $\Phi_N$ used by the one-zone approximation.
The local density at each shell,
\begin{equation}
\rho_{\rm in}(r_i) = \bar\rho(a)\big[1+\delta(r_i,a)\big],
\end{equation}
is used to evaluate the local equilibrium field value $f_{R,\rm min}(\rho_i)$,
and hence the thin-shell parameter is computed \emph{locally} rather than
once for the whole halo:
\begin{equation}
\frac{\Delta R_{th}(r_i)}{R_i} = \frac{\bar f_R(a) - f_{R,\rm min}(\rho_i)}{6\,\Phi_N(r_i)},
\end{equation}
\begin{equation}
S(r_i) = \min\!\left(1,\; 3\,\frac{\Delta R_{th}(r_i)}{R_i}\right).
\end{equation}

Because $S$ now varies shell by shell, the modification factor entering the
source term is likewise evaluated locally, $\epsilon(k,a) \to \epsilon(k,a)\,S(r_i)$,
and the gravitational source term itself becomes shell-dependent:
\begin{equation}
\mathcal{G}(r_i) = \frac{1}{2\pi^2}\int_0^\infty dk\, k^2\, \big[1+S(r_i)\,\epsilon(k,a)\big]\,
\delta(k,a)\,\frac{\sin(kr_i)}{kr_i}.
\label{eq:D17}
\end{equation}

Equation (1)/\ref{eq:evol} is then evaluated explicitly at each shell as

\begin{equation} \label{eq:mast}
\begin{aligned}
&\delta''(r_i) + \left(\frac{3}{a}+\frac{E'}{E}\right)\delta'(r_i) -\frac{4}{3}\frac{\delta'(r_i)^2}{1+\delta(r_i)} = \\
&\frac{3(1+\delta(r_i))\Omega_{m,0}}{2a^5E^2} \left[\mathcal{G}(r_i) - \alpha\,\delta(r_i)\right]
\end{aligned}
\end{equation}

i.e.\ the same master equation, but with the \emph{local}, shell-dependent
screening kernel $\mathcal{G}(r_i)$ of Eq.~\eqref{eq:D17} in place of the
single global $\mathcal{G}$ used in the one-zone approximation. Unlike the
one-zone case, this makes the collapse equation genuinely non-local across
shells: the density profile at every radius feeds back, through $M(<r_i)$
and $\Phi_N(r_i)$, into the effective gravity felt at every other radius.

\section{Conditions for a Nearly Spherical Collapse} \label{sec:condit}

Large values of $\sigma$, and $\omega$ give rise to a non spherical collapse. A large shear will deform the structure, transforming it into an ellipsoid whose axes will collapse at different times. A large vorticity will strongly slow down the collapse. To have a nearly spherical collapse, the two quantities must be compared with the isotropic contraction rate of the perturbation.

Following the covariant characterization of anisotropy (\citep{Ellis1999}), a practical condition for the collapse to remain nearly spherical is to use 
\begin{equation}
\frac{\sqrt{2\sigma^2}}{|\theta|}
\ll
1,
\qquad
\frac{\sqrt{2\omega^2}}{|\theta|}
\ll
1.
\end{equation}
where $\theta$ is the expansion (or contraction scalar). Typical values are 

\begin{align}
\frac{\sqrt{2\sigma^2}}{|\theta|}
&\lesssim
0.03-0.05,
\\
\frac{\sqrt{2\sigma^2}}{|\theta|}
&\lesssim
0.1,
\\
\frac{\sqrt{2\sigma^2}}{|\theta|}
&\gtrsim
0.2-0.3,
\end{align}
\citep{Bertschinger1994,Libenskind2013} 
corresponding respectively to very small deformations, a nearly spherical regime, and a regime where the spherical approximation becomes unreliable.

Another way is using Raychaudhuri equation in the form 

\begin{equation}
\dot\theta
+
\frac{\theta^2}{3}
+
2(\sigma^2-\omega^2)
=
-4\pi G\rho+\cdots .
\end{equation}

The anisotropic stresses enter the Raychaudhuri equation through the combination $(\sigma^2-\omega^2)$. Perturbative and model-based estimates \citep{DelPopolo2013a,DelPopolo2013b,Pace2010,Pace2014,Pace2018}
indicate that this term amounts to a few percent of the gravitational source term, motivating the practical requirement

\begin{equation} \label{eq:condit}
\frac{|\sigma^2-\omega^2|} 
{4\pi G\bar\rho\,\delta}
=
\frac{|\sigma^2-\omega^2|} 
{\frac{3}{2}\,\Omega_m H^2\,\delta}
\lesssim
0.05-0.1.
\end{equation}

In the extended spherical-collapse model,

\begin{equation}
\sigma^2-\omega^2
=
-\frac32
\alpha
\Omega_mH^2\delta,
\end{equation}

Inserting this in Eq. \ref{eq:condit}, 
a practical classification is

\begin{center}
\begin{tabular}{cc}
\hline
$\alpha$ & Interpretation\\
\hline
$\alpha<0.05$ & Excellent nearly spherical approximation\\
$0.05\le\alpha<0.07$ & Nearly spherical\\
$0.07\le\alpha<0.10$ & Weak spherical approximation\\
$\alpha\ge0.10$ & Spherical approximation unreliable\\
\hline
\end{tabular}
\end{center}

The plots versus $z$ assume $\alpha(10^{14} h^{-1} M_{\odot})=0.026$, which is an excellent nearly spherical approximation. 
The correction to the average collapse dynamics is approximately $2.6\%$. 
The value of $\alpha(M)$ at $10^{12} h^{-1} M_{\odot}$ is equal to 0.05925, and in the case $\alpha(M)$ at $10^{15} h^{-1} M_{\odot}$ is $\simeq 0.01431$, while is equal to $\simeq 0.00779$ at $6.7 \times 10^{15} h^{-1} M_{\odot}$, in agreement with the values estimated before.

~\\
~\\
~\\

\bibliographystyle{apsrev4-1}
\bibliography{old_MasterBib1_dedup}

\end{document}